\documentclass[twocolumn]{aastex631}

\begin{document}

\title{Low-redshift analogues of cosmic noon galaxies as laboratories for clumpy star formation}

\author{Jorge M. Santos-Junior}
\affiliation{Observatório do Valongo/UFRJ, 
Ladeira do Pedro Antônio, 43 - Centro,
Rio de Janeiro - RJ, 20080-090, Brazil}
\affiliation{Fundação Planetário do Rio de Janeiro, Av. Padre Leonel Franca, 240, Rio de Janeiro - RJ, 22240-000, Brazil}

\author{Thiago S. Gonçalves}
\affiliation{Observatório do Valongo/UFRJ, Ladeira do Pedro Antônio, 43 - Centro, Rio de Janeiro - RJ, 20080-090, Brazil}

\author{Luidhy Santana-Silva}
\affiliation{Centro Brasileiro de Pesquisas F\'{ı}sicas, Rua Dr. Xavier Sigaud 150, CEP 22290-180, Rio de Janeiro, RJ, Brazil \\
}

\author{Arianna Cortesi}
\affiliation{Observatório do Valongo/UFRJ, Ladeira do Pedro Antônio, 43 - Centro, Rio de Janeiro - RJ, 20080-090, Brazil}
I\affiliation{Institute of Physics, Federal University of Rio de Janeiro, Av. Athos da Silveira Ramos 149, Rio de Janeiro, RJ 21941909, Brazil}

\author{Karin Menendez-Delmestre}
\affiliation{Observatório do Valongo/UFRJ, Ladeira do Pedro Antônio, 43 - Centro, Rio de Janeiro - RJ, 20080-090, Brazil}

\author{Amanda E. de Araujo-Carvalho}
\affiliation{Observatório do Valongo/UFRJ, Ladeira do Pedro Antônio, 43 - Centro, Rio de Janeiro - RJ, 20080-090, Brazil}



\begin{abstract}
It has been established that a significant fraction of star formation at high-redshift occurs in clumpy galaxies. The properties of clumps and their formation mechanisms, however, remain highly debated. In this work we analyse a sample of 18 Supercompact Ultraviolet Luminous Galaxies observed with the OSIRIS spectrograph at the Keck Telescope, targeting their Pa$-\alpha$ emission. These galaxies, although at $z\sim 0.1-0.2$, share many similar properties with star-forming galaxies at cosmic noon. We find a total of 84 star-forming clumps with typical sizes of a few hundred parsecs. The star-forming clumps exhibit low values of velocity shear ($\sim$ 12 km/s) and high velocity dispersion ($\sim$ 70 km/s). The dynamical masses of the clumps are typically higher than gas masses inferred from the measured star-formation rates of each clump. We also artificially redshift our data to emulate observations at $z = 2.2$  and allow for a direct comparison with other galaxies at higher redshift. Our results indicate that, due to the effects of clump clustering and low-resolution observations, high-z clumps appear larger at greater cosmological distances. This underscores the importance of using low-redshift observations to anchor studies at earlier epochs. Finally, our results support the idea of growing clump sizes in star-forming galaxies as a function of redshift, although not to scales of kpc as found by other works without the benefits of adaptive optics or gravitational lensing.  

\end{abstract}


\section{Introduction} \label{sec:intro}

 At cosmic noon ($z\sim 2$), during the peak of star formation in the Universe, star-forming galaxies are characterized by high star-formation rates (SFR) for a given stellar mass and irregular morphologies \citep[e.g.,][]{Law2007, Whitaker2012, Conselice2022}. These galaxies are typically composed of small nodules of star formation, or clumps, with estimated stellar masses of $\sim 10^{7-9}$ M$_{\odot}$ \citep{Guo2015b, soto2017}, a typical SFR of tens to hundreds of  M$_{\odot}$ yr$^{-1}$ \citep{Guo2012} and clump ages ranging from 10$^{6}$ to 10$^{10}$ years\citep{soto2017}. In the local universe, these structures are often observed in irregular dwarf galaxies \citep{elmegreen2009a, elmegreen2021, Bournaud&elmegreen2009}. However, at high redshifts, they are commonly found in  galaxies with gas-rich turbulent disks. \citep{soto2017}. The definition of ``clump'' in the literature can differ depending on the work, ranging from nodules with a luminosity greater than a small percentage ($\sim$8\%) of the overall ultraviolet luminosity of the host galaxy to small clusters of hot stars \citep{Boada2015,Guo2015a}.
 

\cite{Somerville2001} suggest that several occurrences of coalescence over time are the cause of irregular morphologies of galaxies and, therefore, of the presence of star-forming clumps. However, \cite{Hopkins2010} argue that galaxies exhibiting clump alignments (clumpy galaxies) are much more abundant than the expected rate of coalescence, and this process is not efficient enough to explain the observed amount of nodules. The clumps could then arise instead from gas-rich disk fragmentation caused by gravitational instabilities \citep{Toomre1964, Noguchi1999, Immeli2004, Bournaud2007, Bournaud2008,behrendt2015}. In this model, the self-gravitation of stars and gas in a region overcomes the shear effect caused by differential rotation and the dispersion of velocity of stars and gas in the interstellar medium.

These violent disk instability models predict that such clumps would occur frequently in disks with high gas fractions. Indeed,  approximately 50 percent of the baryonic mass in high-redshift star-forming galaxies is composed of gas \citep{TAcconi2010}. This  can cause a fragmentation of the galactic disk due to strong instabilities, allowing the formation of giant clumps with masses in the range of 10$^8$ to 10$^9$ M$_{\odot}$ rotating with the host galaxy \citep{Bournaud2014}.

Also according to \cite{Bournaud2014}, violent instability can drive the formation and growth of bulges through the migration of these nodules to the central regions of the galaxy, which coalesce into a larger clump, building the central bulge of the galaxy. However, in addition to the loss of surface brightness due to the cosmic dimming, the impossibility of obtaining images with high enough angular resolution for the study of individual clump structures requires the use of alternatives that enable the proper characterization of these structures.

Many authors have used the magnification from strong lensing to show that it is possible to observe clumps in detail with a spatial resolution down to 30 pc \citep{Cava2018,Dessauges2017a, Dessauges2017b, Dessauges2018, Dessauges2019, Mestric2022}. 
The commissioning of the JWST has been revolutionary, as it has significantly increased the number of gravitational lensing studies. Consequently, there has been a substantial rise in the number of clumpy galaxies analyzed, enabling the identification of clumps with sub-parsec-scale resolution. This advancement allows for the direct observation of individual star clusters, as can be observed in \cite{ Mowla2022, Mowla2024,Adamo2024, Vanzella2023}. \citet{Claeyssens2025}, in particular, observed a sample of 1956 clumps lensed by the galaxy cluster Abell 2744 with JWST/NIRCam, with estimated R$_{eff}$ between 10 and 700 pc, in 476 galaxies located in the redshift range of 0.7 to 10. Although the increase in statistics is dramatic, the data are still limited to photometric observations.


Numerical simulations have proven to be fundamental in understanding the formation and evolution of clumps in gas-rich galaxies. Recent studies, such as Renaud et al. (2024), demonstrate that the dynamics of the interstellar medium, including turbulence, tidal forces, and shear, significantly influence the formation of these structures. Using simulations of disk galaxies with varying gas fractions, the authors showed that while the scaling relations of clumps follow universal trends, the scatter in these relations increases with the gas fraction, allowing for the presence of massive and highly turbulent clouds. Furthermore, the star formation rate within clumps is more strongly correlated with the excess mass rather than the total mass, suggesting that the evolution of these structures may be modulated by variations in local dynamic conditions.

The VELA simulation series, analyzed by Ceverino et al. (2023), provides a complementary perspective by assessing the impact of stellar feedback on the survival of clumps. The authors investigated different feedback models, with and without the inclusion of kinetic impulses from expanding supernova shells and stellar winds. Their results indicate that massive clumps in disks with high surface density can survive for several dynamical timescales, regardless of feedback strength. However, older and more compact clumps, with low star formation rates, are rapidly dispersed when feedback is stronger. This suggests that the maintenance of these structures depends on the balance between gravitational collapse and the disruptive effects of feedback, a critical factor in understanding galaxy evolution in the early universe.

Finally, the study by Renaud et al. (2024) highlights that the detection and characterization of clumps in simulations rely on structural methods that identify gas overdensities consistently across different conditions. The 12 pc resolution used in their simulations enables the capture of clump morphology and kinematics, although the internal structure of these regions is not fully resolved. The hierarchy of gaseous structures suggests that clumps in gas-rich galaxies may represent scaled-up versions of the giant molecular clouds observed in the Milky Way, but with distinct characteristics due to the extreme conditions of the interstellar medium at high redshifts. These studies emphasize the need for detailed comparisons between simulations and observations to determine to what extent the physical regimes governing star formation in nearby galaxies can be extrapolated to the young universe.

In that context, a plausible alternative for the study of clumps in turbulent disks is the use of low-redshift analogues of cosmic noon star-forming galaxies. In this work we focus particularly on Lyman-Break Analog galaxies (LBAs). These objects at $z\sim 0.1-0.2$ are typically compact, with star formation rates of a few tens of $M_\odot$ yr$^{-1}$ and low extinction values, sharing many physical properties with Lyman break galaxies and other main- sequence star-forming galaxy samples at cosmic noon, being selected to that specific end \citep[see also Section \ref{sec:sample}]{Hoopes2007}. Thus, LBAs are excellent laboratories to study the detailed properties of star formation in galaxies with similar condition to those at high redshift, with the advantage of substantially better angular resolution data and higher signal-to-noise ratio in resolved structures. 

In this work, we investigate the physical properties of star-forming clumps of LBAs, including their gas kinematics and dynamics. Being able to spatially resolve individual clumps, we are able to provide better diagnostics of their physical properties such as size and dynamical support. To explore how the presence and properties of these clumps at earlier times, we artificially redshift the image cubes to $z\sim 2.2$. This allows for a direct investigation of what the observed high-z clumps may comprise in term of internal substructures (e.g., clusters of small clumps).

This article is divided as follows: in Section 2 we describe the LBA sample and observational data; Section 3 describes the methodology to identify clumps in LBAs; Section 4 presents quantitative results and in section 5 we discuss the implications of our measurements in the context of the current knowlegde of star-forming clumps as a function of cosmic time. We present a summary of our findings in Section 6.

The standard cosmological model, with H$_0$= 70 km.s$^{-1}$ Mpc$^{-1}$, $\Omega_m= 0.30$ and $\Omega_{\triangle}= 0.70$, was adopted for all analyses and calculations in this work.

 \section{Sample $\&$ Data}\label{sec:sample}

The ultraviolet luminous galaxy (UVLG) sample was initially defined by searching for objects with far-ultraviolet (FUV) luminosity greater than L$_{FUV} > 2\times$10$^{10}$ L$_{\odot}$ \citep{Heckman2005a}, from Galaxy Evolution Explorer (GALEX) observations overlapping with the SDSS sample \citep{York2000, Martin2005}. Subsequently, \citet{Hoopes2007} defined a subsample with FUV surface brightness L$_{UV} > 10^9$ L$_{\odot}$ kpc$^{-2}$, in an attempt to mimic actual cosmic noon star-forming galaxies such as the Lyman-break galaxies (LBGs) -- hence their denomination as {\it Lyman-break analogs} (LBAs).

Later, many authors have shown that LBAs indeed reproduce many of the observed properties of $z\sim 2$ star-forming galaxies such as stellar masses, SFR, mass-metallicity relations \citep{Heckman2005a,Hoopes2007}, dust extinction \citep{Overzier2011}, ionized gas kinematics \citep{Basu-Zych2009,Thiago2010}, radio and X-ray luminosities \citep{Basu-Zych2007,Basu-Zych2013}, molecular gas reservoirs \citep{Thiago2014, contursi2017}, and ionization parameters \citep{Loaiza-Agudelo2020}.

\citet{Overzier2009,Overzier2010}, in particular, have used Hubble Space Telescope (HST) observations carried out in the ultraviolet and optical wavelengths to reveal morphologies and structural features typical of starburst galaxies observed in the distant universe, such as massive star-forming clumps and evidence for outflows, as well as coalescence interactions (mergers). Their work, however, emphasized that many of the characteristic features of merger activity was lost due to observational biases. This is also true for the ionized gas kinematics, for which high-redshift observations render galaxy smoother and more similar to regularly rotating disks \citep{Thiago2010, Hunh2016}.

The sample used in this work comprises 18 LBAs, (presented in Table \ref{tab:thiago1}), with typical redshifts between $0.1 \lesssim z \lesssim 0.25$, observed using the OSIRIS integral field unit (IFU) \citep{Larkin2006}, attached to the Keck telescope ($R\sim 3000$), located on Mount Mauna Kea, Hawaii. We have used the instrument with plate scales of 50 mas per pixel in most cases, for a field-of-view (FOV) of approximately 2'' x 3''. For the larger objects in the sample, we have used the 100 mas plate scale, with double the linear dimensions and a much larger FOV. We have targeted the Pa-$\alpha$ emission line (1875.1 nm rest wavelength), which falls between the Kn3 and Kn5 bands depending on the exact galaxy redshift. The final angular resolution, as measured by the PSF of a point source in the same configuration observed right before the galaxy, is approximately 100 mas in most cases, corresponding to $\sim 200$ pc for typical LBA redshifts. For more details on the observations, we refer the reader to \citet{Thiago2010}.


    

\begin{table*}
\caption{Properties of our LBAs sample}
    \centering
    \begin{tabular}{c|c|c|c|c|c|C}
\hline
\hline
ID&z&Spaxel&AOFWHM&SFR&R$_{eff}$&log M \\

  & &Scale(mas)&(mas)&[M$_\odot$.y$^{-1}$]&[kpc]& [M$_\odot$] \\
\hline
005527&0.167&50&90&22.7&0.36&9.70\\
015028&0.147&50&82&19.4&1.34&10.3\\
021348&0.219&100&177&6.7&0.38&10.5\\
032845&0.142&50&103&6.7&0.86&9.8\\
035733&0.204&100&116&9.6&1.00&10\\
040208&0.139&50&80&2.0&0.80&9.5\\
080844&0.096&100&187&3.7&0.08&9.8\\
082001&0.218&50&69&20.3&2.78&9.8\\
083803&0.143&50&105&4.5&1.02&9.5\\
092600&0.181&50&101&13.2&0.68&9.1\\
093813&0.107&50&77&13.0&0.65&9.4\\
101211&0.246&50&96&4.3&N/A&9.8\\
113303&0.241&50&76&5.3&1.36&9.1\\
135355&0.199&50&68&17.1&1.45&9.9\\
143417&0.180&50&98&25.4&0.90&10.7\\
210358&0.137&50&65&41.3&0.44&10.9\\
214500&0.204&50&70&13.6&1.13&9.9\\
231812&0.252&100&130&30.8&2.77&10.0\\
\hline
    \end{tabular}
    \renewcommand{\footnotesize}{\scriptsize}

 \footnote{List of properties of all LBA galaxies analyzed in this work. Columns show, in order, galaxy ID (as presented in \cite{Overzier2009}), redshift, original spaxel scale with the OSIRIS instrument, AO-assisted FWHM of the observations, and the physical properties as previously determined by \cite{Hoopes2007}: star-formation rates, effective radii and stellar mass.}


    \label{tab:thiago1}
\end{table*}




The 18 datacubes were subjected to the data reduction routine of the OSIRIS equipment \citep{Wright2009} and a set of pipelines developed by the authors. A routine was developed to collapse the spatial dimensions of each cube and generate a single spectrum for the entire image. A Gaussian fit was then performed on the spectrum to determine the central wavelength of the Pa$-\alpha$ line and the Gaussian fit to the emission, with negligible contribution from the NIR continuum. The total Pa$-\alpha$ flux of the galaxy was obtained using the interval of three times the dispersion value (3$\sigma$) around the peak of the fit. Once the central wavelength and line width were determined, all frames between -3$\sigma$ and +3$\sigma$ were collapsed along the spectral dimension to produce the two-dimensional images of the Pa$-\alpha$ distribution in the galaxies.

Additionally, first- and second-order moment maps (velocity and velocity dispersion, respectively), were produced by fitting Gaussian profiles to spectra in each individual spaxel. Only those fits above $S/N=6$ were used for the kinematic analysis, with the remaining spaxels being discarded.



\subsection{Simulated observations at high-z}

In addition to the data described above, we produced simulated datacubes as though LBAs were observed at $z=2.2$, again with the aid of adaptive optics. This specific redshift was chosen as to reproduce typical distances of cosmic noon galaxies, while at the same time avoiding strong sky emission lines and permitting clean detections of the H$\alpha$ line. This follows the same OSIRIS IFU prescriptions and reduction pipeline described in \citet{Law2006} and \citet{Law2009}.

The datacubes were spatially smoothed using the same typical angular resolution and spaxel scale of OSIRIS, assuming an angular diameter distance at higher redshift. Fluxes were also corrected using the integrated H$\alpha$ fluxes from SDSS, which are intrinsically more luminous, but will suffer stronger dimming due to the larger luminosity distances involved.

For further information, we refer the reader to \citet{Thiago2010}.


\section{Clump Identification}

Star-forming clumps are regions of high gas and dust density within star-forming galaxies, where intense star formation processes occur. These regions are characterized by a significantly higher concentration of matter compared to the surrounding environment, facilitating the collision and agglomeration of material to form stars. The precise definition of clumps may vary depending on the criteria used, including surface density, star formation rate, and physical size. 

 According to \cite{Boada2015} individual clumps are discerned based on the ultraviolet (UV) luminosity disparity between each clump and its flux relative to the average emitted by the surrounding environment of the host galaxy. The authors define clumps as having UV luminosity surpassing 1$\%$ of the cumulative UV emission of the galaxy (L$_{\rm clump}$ / L$_{\rm galaxy} > 0.01$) . On the other hand, \cite{Guo2015a} adopt a threshold of UV luminosity exceeding 8$\%$ for clump selection.




\begin{table}
\caption{Fellwalker configuration parameters values}
    \centering
\begin{tabular}{l|c}
\hline
\hline

  FELLWALKER.ALLOWDEGE= &0 \\
  FELLWALKER.CLEANITER=&3 \\
  FELLWALKER.FLATSLOPE=& 2.0*RMS \\
  FELLWALKER.FWHMBEAM=&2  \\
  FELLWALKER.MAXBAD=&0.05  \\
  FELLWALKER.MAXJUMP=&4  \\
  FELLWALKER.MINDIP=& 2.0*RMS \\
   FELLWALKER.MINHEIGHT=& 2.0*RMS\\
   FELLWALKER.MINPIX=&4  \\
   FELLWALKER.NOISE=&2.0*RMS \\
   FELLWALKER.RMS=& from image  \\

\hline
    \end{tabular}
    
    \label{tab:fell_par1}
\end{table}

In this work we implemented a two-step approach for the selection of clumps. First, we used the CUPID package, which employs the FellWalker algorithm \citep{Berry2015} to identify clumps. This has been extensively used in the literature to search for cores in Galactic molecular clouds and star-forming regions \citep[e.g.,][]{Rigby2016,ZHANG2018,Liu2020}, and has been shown to perform as well as or better than similar clump-finding algorithms \citep{Li2020}. Furthermore, it has been used recently to identify clumps in other galaxies \citep{Spilker2022,Posses2024}. We therefore choose FellWalker to select and identify clumps in in each of the Paschen$-{\alpha}$ (Pa$-\alpha)$ flux maps of LBAs.


\begin{figure*}
\begin{center}
\includegraphics [width=\linewidth]
{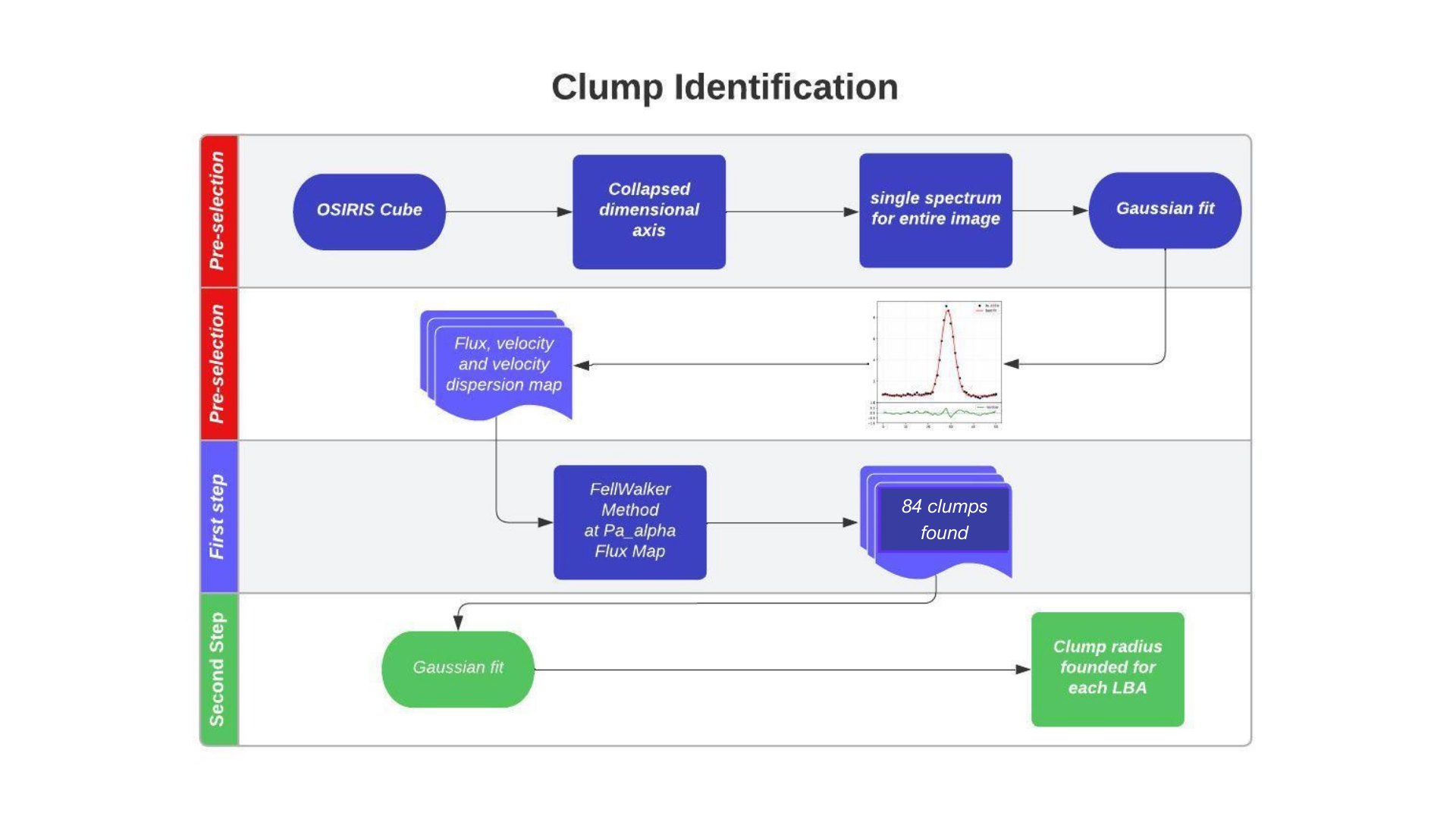}
\caption{This diagram summarizes the steps used to selecting the clumps in our sample. We first collapse the datacubes along the spatial dimensions to obtain a one-dimensional spectrum for the entire galaxy as a function of wavelength. Next we fit a Gaussian profile to the Pa$-\alpha$ line to select the wavelength channels collapsed to produce two-dimensional maps of flux for each object. Later we apply the Fellwalker method to the flux maps to identify the clumps, as described in the text. Finally, we fit a Gaussian to the angular projection of each clump, crossing its center, to determine the clump radii for each of the LBAs. A total of 84 clumps were identified. Subsequently, a test was performed to determine whether each clump is resolved or not, resulting in 38 clumps that could be resolved, according to the criterion described in Equation \ref{eq:resolution}, which will be analyzed throughout this paper.}
\label{fig:Fluxograma}
\end{center}
\end{figure*}

For each pixel in the flux map, we apply the FellWalker algorithm to select the nearest flux peak. Starting from a pixel, this algorithm searches for the adjacent pixel with the highest count value. The procedure is repeated until no higher value is found, thus creating a maximum gradient route towards a local peak. The routine then searches for pixels with flux values greater than the last pixel (preliminary-peak) within a 4-pixel radius, in all directions. If no pixels satisfying this condition are found, the last pixel searched is considered as the peak of the clump. The value of 4 (denoted MAXJUMP in the input parameters) was adopted because it is approximately twice the AO-FWHM value for each observed image. All other pixels that end the search at this peak and those that were used in the route are considered as belonging to the clump.

Slopes flatter than a threshold determined below a given S/N level as an input parameter are discarded, in order to exclude flatter features in galaxies. Therefore, only high-gradient or bright peaks are selected as clumps. Pixels with counts below the NOISE threshold are also discarded.





The noise value used to identify where clumps would not be searched was obtained by calculating the standard deviation of pixel values in velocity channels blueward and redward of the integrated emission line, beyond 5$\sigma$ of the velocity dispersion. We have verified that continuum emission is negligible in most cases and does not affect our measurements. Any pixels below twice this root mean square (RMS) value would be discarded. We also take into account the angular resolution of the data, with a minimum of 4 pixels belonging to the clump as a requirement to proceed to the next step. 
The FellWalker's parameters used in the clump search are listed in Table \ref{tab:fell_par1}.

\begin{figure*}
\begin{center}
\includegraphics [width=\linewidth]
{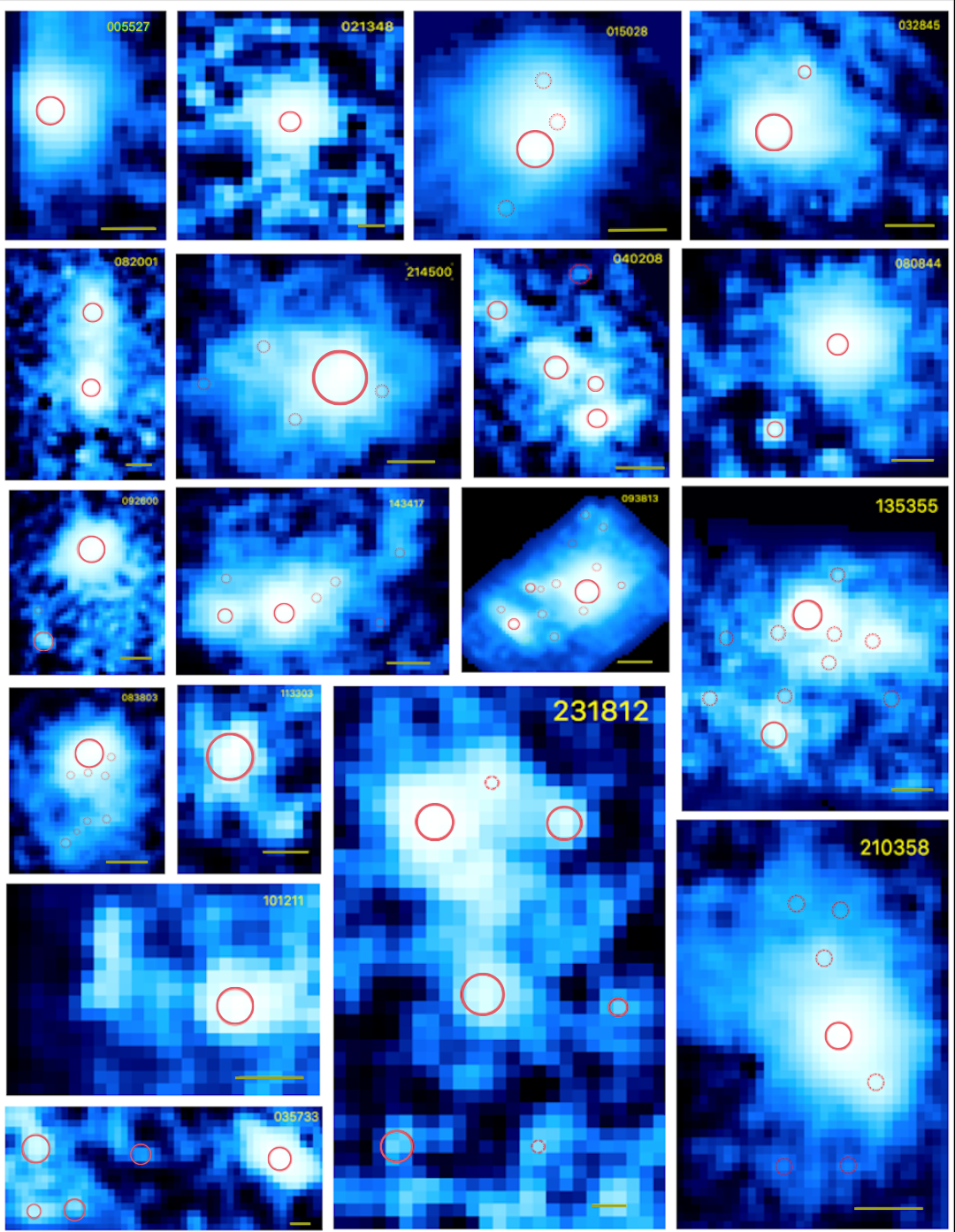}
\caption{LBAs with their respective clumps identified. The red circles indicate the positions of the 84 identified clumps. The solid red circles show the 38 clumps with resolution above the threshold presented in Equation 1, while the dotted red circles denote the remaining unresolved clumps. The yellow line in the lower left corner represents a physical scale of 1 kpc.}
\label{fig:lbas_clumps}
\end{center}
\end{figure*}

This method was applied to each of the 18 Pa$-\alpha$ flux maps of the sample, and is summarized in Figure \ref{fig:Fluxograma}. This method results in a file with an identification number in a catalog with several clump information, such as the position of the peaks on the x and y axis, the central position of the clump on both axis, the radius found on both axes, in addition to the counts on the peaks and the number of pixels in each one of the identified clumps. We identify 84 clump structures in our galaxy sample. These clumps are shown in Figure \ref{fig:lbas_clumps}. By definition, each galaxy will have a minimum of 1 clump.

We then perform a gaussian fit to the count profiles along the x- and y-axes of the data, centered at the positiong of the clumps determined by FellWalker. This is done to determine clump sizes, since those quantities are not reliably measured in the previous package. The clump radius is then defined as the quadrature average of the FWHM along each axis, i.e., $R=1/2 \sqrt{{\rm FWHM}_x\times {\rm FWHM}_y}$ 


\label{fig:lbas_clumps1}

\begin{table*} 
\caption{Properties of resolved star-forming clumps}
\centering
 \begin{tabular}{c|c|c|c|c|c|c|c|c}
\hline
\hline
ID    & R & V$_s$ & V$_{err}$ & $\sigma$  & $\sigma_{err}$ & SFR & log M$_{gas}$ & log M$_{dyn}$\\
  & [pc] & [km.s$^{-1}$] & [km.s$^{-1}$] & [km.s$^{-1}$] & [km.$s^{-1}$] & [M$_\odot$.y$^{-1}$ ]& [M$_\odot$]& [M$_\odot$] \\

\hline
005527c1  & 247.08         & 4.53  & 0.74 & 92.26  & 1.07  & 7.84  & 9.01  & 9.09    \\
015028c1  & 352.19         & 20.74 & 1.55 & 88.47  & 2.41  & 2.82  & 8.78  & 9.20    \\
015028c2  & \textless 300  & N/A   & N/A  & N/A    & N/A   & N/A   & N/A   & N/A     \\
015028c3  & \textless 300  & N/A   & N/A  & N/A    & N/A   & N/A   & N/A   & N/A     \\
015028c4  & \textless 300  & N/A   & N/A  & N/A    & N/A   & N/A   & N/A   & N/A     \\
021348c1  & 407.09         & 1.20  & 1.12 & 74.87  & 1.84  & 2.08  & 8.72  & 9.12    \\
032845c1  & 387.06         & 11.60 & 1.59 & 64.01  & 2.56  & 0.95  & 8.46  & 8.96    \\
032845c2  & 124.86         & 14.28 & 1.96 & 52.11  & 0.76  & 0.05  & 7.27  & 8.29    \\
035733c1  & 576.39         & 15.36 & 2.92 & 72.7   & 4.75  & 2.26  & 8.83  & 9.25    \\
035733c2  & 579.74         & 5.67  & 4.00 & 52.9   & 6.53  & 2.58  & 8.87  & 8.97    \\
035733c3  & 392.08         & N/A   & N/A  & N/A    & N/A & 0.09    & N/A   & N/A     \\
035733c4  & 462.45         & N/A   & N/A  & N/A    & N/A  & 0.05   & N/A   & N/A     \\
035733c5  & 512.72         & N/A   & N/A  & N/A    &  N/A  & 0.03   & N/A   & N/A     \\
040208c1  & 295.54         & 7.93  & 5.69 & 33.76  & 11.63 & 0.09  & 7.67  & 8.29    \\
040208c2  & 215.83         & 9.76  & 9.18 & 28.99  & 20.63 & 0.06  & 7.46  & 8.02    \\
040208c3  & 155.74         & 6.89  & 4.86 & 39.57  & 10.40 & 0.02  & 7.04  & 8.15    \\
040208c4  & 168.01         & 3.32  & 2.57 & 25.56  & 5.42  & 0.01  & 6.84  & 7.80    \\
040208c5  & 208.47         & N/A   & N/A  & N/A    & 0.01  & N/A   & N/A   & N/A     \\
080844c1  & 250.79         & 2.31  & 0.60 & 93.84  & 1.07  & 0.59  & 8.21  & 9.11    \\
080844c2  & 192.10         & 1.62  & 1.34 & 42.25  & 2.59  & 0.06  & 7.43  & 8.30    \\
082001c1  & 303.37         & 10.11 & 0.84 & 61.84  & 1.50  & 1.39  & 8.52  & 8.83    \\
082001c2  & 444.47         & 20.89 & 2.32 & 54.45  & 4.10  & 0.62  & 8.37  & 8.88    \\
083803c1  & 447.13         & 9.38  & 1.62 & 36.49  & 3.56  & 0.68  & 8.40  & 8.54    \\
083803c2  & \textless 400  & N/A   & N/A  & N/A    & N/A   & N/A   & N/A   & N/A     \\
083803c3  & \textless 400  & N/A   & N/A  & N/A    & N/A   & N/A   & N/A   & N/A     \\
083803c4  & \textless 400  & N/A   & N/A  & N/A    & N/A   & N/A   & N/A   & N/A     \\
083803c5  & \textless 400  & N/A   & N/A  & N/A    & N/A   & N/A   & N/A   & N/A     \\
083803c6  & \textless 400  & N/A   & N/A  & N/A    & N/A   & N/A   & N/A   & N/A     \\
083803c7  & \textless 400  & N/A   & N/A  & N/A    & N/A   & N/A   & N/A   & N/A     \\
083803c8  & \textless 400  & N/A   & N/A  & N/A    & N/A   & N/A   & N/A   & N/A     \\
083803c9  & \textless 400  & N/A   & N/A  & N/A    & N/A   & N/A   & N/A   & N/A     \\
092600c1  & 518.26         & 16.86 & 1.60 & 79.00  & 2.58  & 3.41  & 8.93  & 9.27    \\
092600c2  & 283.52         & 6.30  & 6.04 & 49.92  & 10.93 & 0.16  & 7.83  & 8.61    \\
093813c1  & 329.86         & 14.27 & 0.49 & 77.81  & 0.72  & 1.48  & 8.56  & 9.06    \\
093813c2  & 158.57         & 2.60  & 0.41 & 54.42  & 0.70  & 0.39  & 7.97  & 8.44    \\
093813c3  & 119.42         & 3.23  & 1.31 & 48.03  & 2.57  & 0.04  & 7.19  & 8.20    \\
093813c4  & \textless 100  & N/A   & N/A  & N/A    & N/A   & N/A   & N/A   & N/A     \\
093813c5  & \textless 100  & N/A   & N/A  & N/A    & N/A   & N/A   & N/A   & N/A     \\
093813c6  & \textless 100  & N/A   & N/A  & N/A    & N/A   & N/A   & N/A   & N/A     \\
093813c7  & \textless 100  & N/A   & N/A  & N/A    & N/A   & N/A   & N/A   & N/A     \\
093813c8  & \textless 100  & N/A   & N/A  & N/A    & N/A   & N/A   & N/A   & N/A     \\
093813c9  & \textless 100  & N/A   & N/A  & N/A    & N/A   & N/A   & N/A   & N/A     \\
093813c10 & \textless 100  & N/A   & N/A  & N/A    & N/A   & N/A   & N/A   & N/A     \\
093813c11 & \textless 100  & N/A   & N/A  & N/A    & N/A   & N/A   & N/A   & N/A     \\
093813c12 & \textless 100  & N/A   & N/A  & N/A    & N/A   & N/A   & N/A   & N/A     \\
093813c13 & \textless 100  & N/A   & N/A  & N/A    & N/A   & N/A   & N/A   & N/A     \\
093813c14 & \textless 100  & N/A   & N/A  & N/A    & N/A   & N/A   & N/A   & N/A     \\

 \end{tabular}
    \label{tab:all_clumps1}
\end{table*}

\begin{table*} 
\setcounter{table}{2} 
\caption{Properties of star-forming clumps -- Continued}
\centering
 \begin{tabular}{c|c|c|c|c|c|c|c|c}
\hline
\hline
ID    & R & V$_s$ & V$_{err}$ & $\sigma$  & $\sigma_{err}$ & SFR & log M$_{gas}$ & log M$_{dyn}$\\
  & [pc] & [km.s$^{-1}$] & [km.s$^{-1}$] & [km.s$^{-1}$] & [km.$s^{-1}$] & [M$_\odot$.y$^{-1}$ ]& [M$_\odot$]& [M$_\odot$] \\

\hline

101211c1  & 334.22         & 10.62 & 2.64 & 62.35  & 20.78  &  0.34 &  8.11 &  8.88  \\
113303c1  & 384.34         & 6.24  & 4.55 & 24.11  & 0.19  & 2.66  & 8.78  & 8.11    \\
135355c1  & 358.25         & 11.78 & 4.74 & 62.71  & 7.34  & 0.60  & 8.30  & 8.91    \\
135355c2  & 251.43         & 7.69  & 5.60 & 98.32  & 22.58 & 0.10  & 7.66  & 9.15    \\
135355c3  & \textless 200  & N/A   & N/A  & N/A    & N/A   & N/A   & N/A   & N/A     \\
135355c4  & \textless 200  & N/A   & N/A  & N/A    & N/A   & N/A   & N/A   & N/A     \\
135355c5  & \textless 200  & N/A   & N/A  & N/A    & N/A   & N/A   & N/A   & N/A     \\
135355c6  & \textless 200  & N/A   & N/A  & N/A    & N/A   & N/A   & N/A   & N/A     \\
135355c7  & \textless 200  & N/A   & N/A  & N/A    & N/A   & N/A   & N/A   & N/A     \\
135355c8  & \textless 200  & N/A   & N/A  & N/A    & N/A   & N/A   & N/A   & N/A     \\
135355c9  & \textless 200  & N/A   & N/A  & N/A    & N/A   & N/A   & N/A   & N/A     \\
135355c10 & \textless 200  & N/A   & N/A  & N/A    & N/A   & N/A   & N/A   & N/A     \\
135355c11 & \textless 200  & N/A   & N/A  & N/A    & N/A   & N/A   & N/A   & N/A     \\
143417c1  & 257.98         & 5.72  & 4.89 & 73.21  & 7.75  & 1.22  & 8.44  & 8.91    \\
143417c2  & 212.45         & 39.41 & 14.07 &  75.57 &  22.61 &  0.90 &  8.12 &  8.85 \\
143417c3  & \textless 200  & N/A   & N/A  & N/A    & N/A   & N/A   & N/A   & N/A     \\
143417c4  & \textless 200  & N/A   & N/A  & N/A    & N/A   & N/A   & N/A   & N/A     \\
143417c5  & \textless 200  & N/A   & N/A  & N/A    & N/A   & N/A   & N/A   & N/A     \\
143417c6  & \textless 200  & N/A   & N/A  & N/A    & N/A   & N/A   & N/A   & N/A     \\
143417c7  & \textless 200  & N/A   & N/A  & N/A    & N/A   & N/A   & N/A   & N/A     \\
210358c1  & 186.55         & 17.59 & 0.60 & 187.15  & 0.98 & 4.22 & 8.74 &  9.58  \\
210358c2  & \textless 150  & N/A   & N/A  & N/A    & N/A   & N/A   & N/A   & N/A     \\
210358c3  & \textless 150  & N/A   & N/A  & N/A    & N/A   & N/A   & N/A   & N/A     \\
210358c4  & \textless 150  & N/A   & N/A  & N/A    & N/A   & N/A   & N/A   & N/A     \\
210358c5  & \textless 150  & N/A   & N/A  & N/A    & N/A   & N/A   & N/A   & N/A     \\
214500c1  & 506.02         & 32.48 & 2.50 & 55.78  & 4.53  & 1.68  & 8.71  & 8.96    \\
214500c2  & \textless 400  & N/A   & N/A  & N/A    & N/A   & N/A   & N/A   & N/A     \\
214500c3  & \textless 400  & N/A   & N/A  & N/A    & N/A   & N/A   & N/A   & N/A     \\
214500c4  & \textless 400  & N/A   & N/A  & N/A    & N/A   & N/A   & N/A   & N/A     \\
214500c5  & \textless 400  & N/A   & N/A  & N/A    & N/A   & N/A   & N/A   & N/A     \\
231812c1  & 542.76         & N/A   & N/A  & N/A    &  N/A & 1.67   & N/A   & N/A     \\
231812c2  & 479.83         & N/A   & N/A  & N/A    & N/A  & 0.35   & N/A   & N/A     \\
231812c3  & 818.08         & N/A   & N/A  & N/A    & N/A  & 0.91   & N/A   & N/A     \\
231812c4  & 428.70         & N/A   & N/A  & N/A    & N/A  & 0.13   & N/A   & N/A     \\
231812c5  & 397.24         & N/A   & N/A  & N/A    &  N/A & 0.14   & N/A   & N/A     \\
231812c6  & \textless{}300 & N/A   & N/A  & N/A    & N/A   & N/A   & N/A   & N/A     \\
231812c7  & \textless{}300 & N/A   & N/A  & N/A    & N/A   & N/A   & N/A   & N/A 

 \end{tabular}

 \footnote{The ID column provides the identification of the 84 clumps found in the 18 LBAs. The R[pc] column contains the radius value, in parsecs, obtained after identifying the clump using the FellWalker method, Gaussian fitting, and comparison with the AO-FWHM of the system. According to our criterion (see equation \ref{eq:resolution}),  38 clumps were considered resolved. Values marked with $<$  represent clumps that are unresolved according to the criteria presented in the text. The V$_{s}$ and V$_{err}$ 
 columns correspond to the shear velocity and its respective errors, both in km $s^{-1}$. The $\sigma$ and $\sigma_{err}$ columns represent the velocity dispersion and its respective error, also in km s$^{-1}$. The Star Formation Rate (SFR), given in solar masses per year, was obtained as described in the text below. Finally, the gas mass was derived using the Schmidt-Kennicutt relation (see equation  \ref{eq:SK}), and the dynamical mass was calculated using equation \ref{eq:dynamic mass}.}
    \label{tab:all_clumps1}
\end{table*}



\subsection{Number of Clumps}

After identifying the clumps in each of the 18 Pa$-\alpha$ flux maps, we obtained a final number of 84 clumps, between 1 and 14 per galaxy. Determining the clump radius is crucial for our analysis because all other parameters depend on this measurement. We tested whether the radius values obtained by the two-step selection process allow for the spatial resolution of clumps. Thus, since the radius value obtained is, for each clump, an average of the adjusted Gaussian dispersions ($\sigma_g$), in order to be able to compare it with the AO-FWHM of the system, we require the measured size to be at least 20\% greater than the PSF FWHM; otherwise, the clump is considered unresolved and the measured size would be an upper limit. This threshold was chosen after careful examination of the clumps, as visual inspection of clumps larger than this typically showed velocity gradients demonstrating its internal structure was at least partially resolved.

\begin{equation}
        {\rm threshold} =  1.2\; \frac{2.355    \sigma_g} { {\rm FWHM}_{\rm PSF} }.
        \label{eq:resolution}      
    \end{equation}

Only 38 clumps met our resolution criterion and were thus considered resolved. For these, velocity and velocity dispersion values were obtained. The Star Formation Rate (SFR), which depends on the integrated flux, was obtained for all clumps. The characteristic values of each clump can be seen in Table \ref{tab:all_clumps1}.

\section{Measuring Clump Properties}


\subsection{Star formation}


We estimate star formation rates (SFR) for each individual clump from the Pa$-\alpha$ flux within that clump. Absolute flux calibration represents a major challenge in AO-assisted IFU data, and thus we take an alternative approach.

We measure the counts within one SDSS fiber photometric aperture (3''), and calibrate that to the measured SFR from H$\alpha$. We assume the SFR-H$\alpha$ ratio to be constant throughout the galaxy, and measure the counts in each clump, assigning individual SFR measurements. In that sense, the quoted SFR values can be interpreted as the ratio between the counts within the clump and those within an aperture equivalent to the SDSS fiber aperture \citep[e.g.,][]{Kewley2005}. We assume that extinction throughout the galaxy, as \citet{Overzier2011} has shown that extinction values in LBAs are typically $E(B - V) \lesssim  0.1$, small enough that this will not strongly affect our results.



The field of view of the Keck/OSIRIS instrument is typically smaller than the diameter of the 3” fiber of the SDSS (with the exception of those observed with the 100 mas plate scale). To correct for that effect, the GALFIT data analysis algorithm \citep{Peng2010} was used to model ellipsoidal profiles of our objects and calculate the expected flux in Pa$_\alpha$ for the required diameter extending beyond the field of view of our instrument. In our modeling approach, we employed an exponential profile (Sérsic = 1). The derived SFRs are shown in Table \ref{tab:all_clumps1}.



\subsection{Velocity shear and velocity dispersion}

\begin{figure}
\begin{center}
\includegraphics[width=\linewidth]{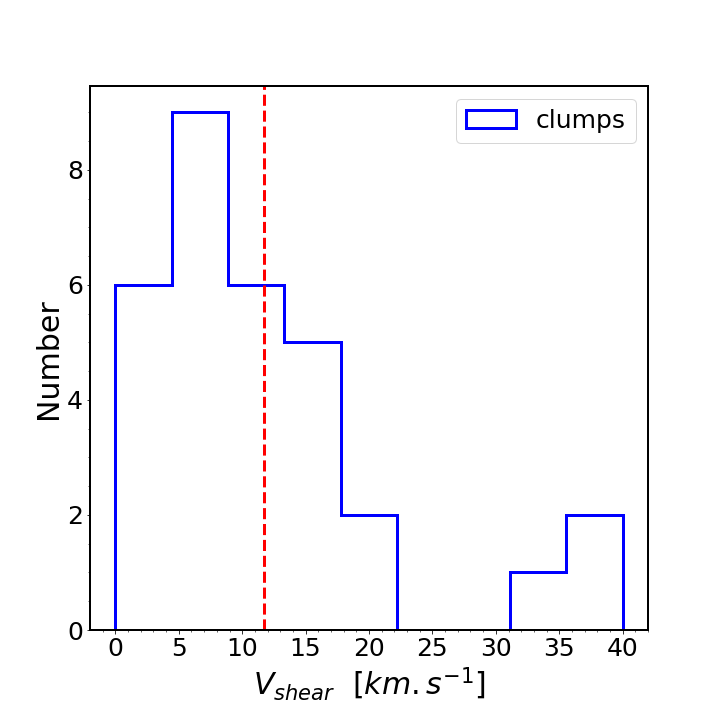}
\caption{Velocity shear distribution for the 31 identified resolved clumps. The red dashed line highlight the mean value for the velocity distribution  ($\sim$ 11 km s$^{-1}$). }
\label{fig:velocity}
\end{center}
\end{figure}

\begin{figure}
\begin{center}
\includegraphics[width=\linewidth]{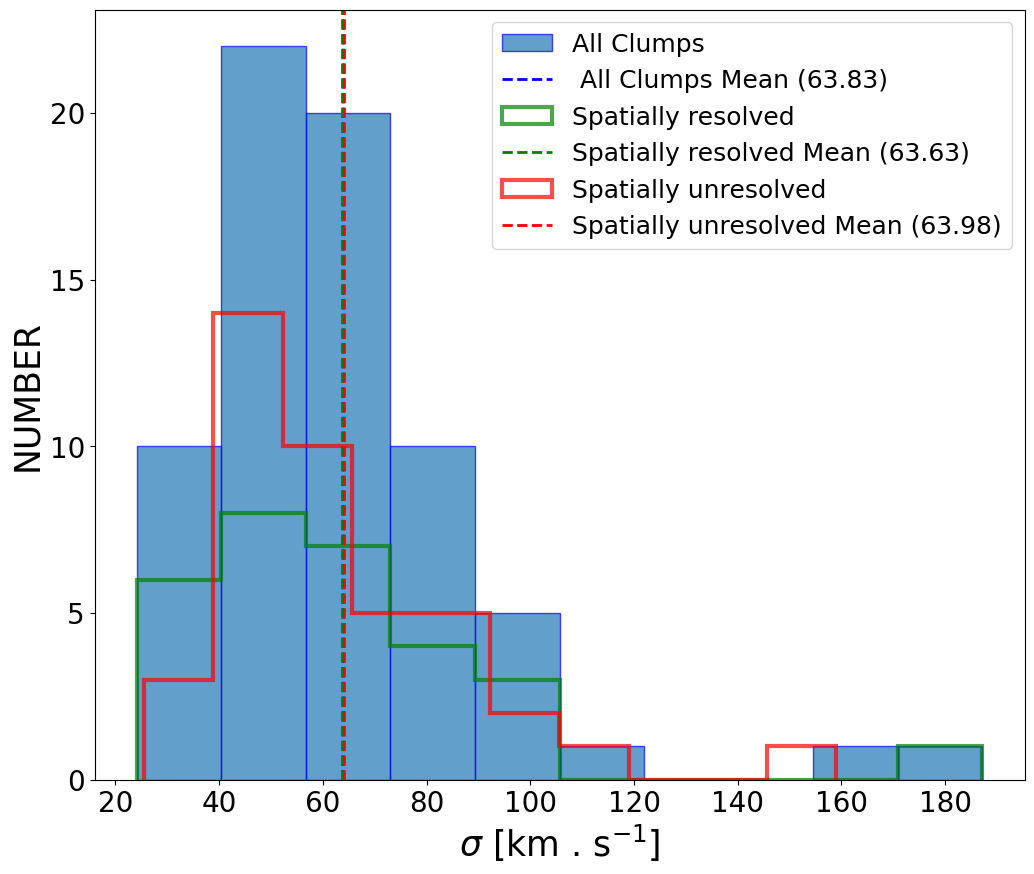}

\caption{Velocity dispersion distribution for the 77 identified clumps. The blue shaded histogram show the distribution for all clumps, while green and red lines represent resolved and unresolved clumps, respectively. The dashed lines highlight the mean values for the velocity dispersion distribution of each sample, with indistiguishable values between them ($\sim$ 64 km s$^{-1}$).} 
\label{fig:sigma}
\end{center}
\end{figure}

To generate the kinematic velocity and velocity dispersion maps, a Gaussian function was fitted to the emission line detected in each spaxel, as described in \cite{Thiago2010}. The peak of the resulting Gaussian fit from the integrated one-dimensional spectrum of the collapsed image cube was used as the zero-velocity restframe of the galaxy. The produced moment maps were limited to a minimum S/N threshold of 6, which represents the detection limit for the surface density of star formation at approximately 0.1 M$_\odot$yr$^{-1}$ kpc$^{-2}$, as explained in \cite{ForsterSchreiber2009} and \cite{Thiago2010}. 


The first moment (velocity) maps were used to obtain velocity shear measurements for each {\it resolved} identified clump -- 46 unresolved clumps are excluded from this stage since such measurements are impossible. We select the regions corresponding to each clump to determine the maximum and minimum velocities, and the velocity shear is expressed by V$_s$ = (V$_{max}$ - V$_{min}$)/2. Figure \ref{fig:velocity} presents the velocity shear distribution for the 31 available clumps, with the mean value highlighted as the dashed red line.

To obtain the velocity dispersion values ($\sigma$) in each clump, the same process described above was used to identify the clumps in the second moment (velocity dispersion) maps. The individual values are equivalent to the means of the velocity dispersion on the clump region. Figure \ref{fig:sigma} presents the dispersion velocity distribution for 77 clumps and the mean for all clumps. This value is typically similar to the average value found for the host galaxy \citep{Thiago2010}. There is no significant distinction for the measured velocity dispersions between resolved and unresolved clumps.

\subsection{Gas content and virial mass}

As shown in \cite{Thiago2014}, LBAs follow the same Schmidt-Kennicutt relationship as local galaxies, albeit at much higher densities:

\begin{equation}
\Sigma_{SFR} = A \Sigma_{gas}^{N},
\label{eq:SK}
\end{equation}
where $\Sigma_{SFR}$ is the star formation rate density, in $M_{\odot}$ yr$^{-1}$ kpc$^{-2}$, and $\Sigma_{gas}$ is the gas density, in $M_{\odot}$ pc$^{-2}$. The constants $A = 2.5 \times 10^{-4}$ and $N = 1.4$ are empirically determined by \cite{Kennicutt1998}. They have a well-defined main sequence, differing from local galaxies only in their efficiency in converting gas into stars. The same Schmidt-Kennicutt relationship was used to estimate the gas mass of the individual clumps, assuming it still holds for individual regions. We are aware that this assumption is likely untrue, especially at high redshift \citep[Gon\c{c}alves et al. in prep]{Freundlich2013, Nagy2023}, potentially affecting the gas mass calculations. In particular, higher SF efficiency in star forming regions would mean smaller gas masses than computed here. However, this effect would not modify gas masses by more than .3 dex, as evidenced by resolved ALMA observations of LBA galaxies (Gon\c{c}alves et al. in prep), which is still typically within our uncertainties.



We have also independently inferred the dynamic masses of the clumps. We assume that the gas kinematics is dominated by random movements within the chosen region, as can be seen by the values of the velocity shear and the velocity dispersion, see Table \ref{tab:all_clumps1} 
 (i.e., $\sigma_{\rm clump} >$ V$_{\rm shear}$). This is likely an upper limit, as the velocity dispersion $\sigma_{\rm clump}$ is probably enhanced by the feedback from stellar winds due to the high star formation rates \citep{Thiago2010}. 

To calculate the dynamic mass, we follow the relation in Equation \ref{eq:dynamic mass}, where $\sigma$ is the velocity dispersion, $R$ is the size of the object and $G$ the gravitational constant. The distribution of mass and the velocity field will be responsible for the value adopted for the C factor. This value can vary between 1 and 5, depending mainly on the mass density profile, the anisotropy of the velocity and the contribution of random movements or rotation \citep{erb2006}, being 1 for disk-like objects and 5 for spherical objects.

In this work, the value of C = 5 was used because we assume that the clumps have spherical symmetry, a reasonable assumption given their $V_S/\sigma$ ratios.
\begin{center}
    \begin{equation}
        M_{dyn} = C\frac{R \sigma^{2}} { G }
        \label{eq:dynamic mass}
    \end{equation}
\end{center}

\begin{center}
\begin{figure}
\includegraphics[width=\linewidth]{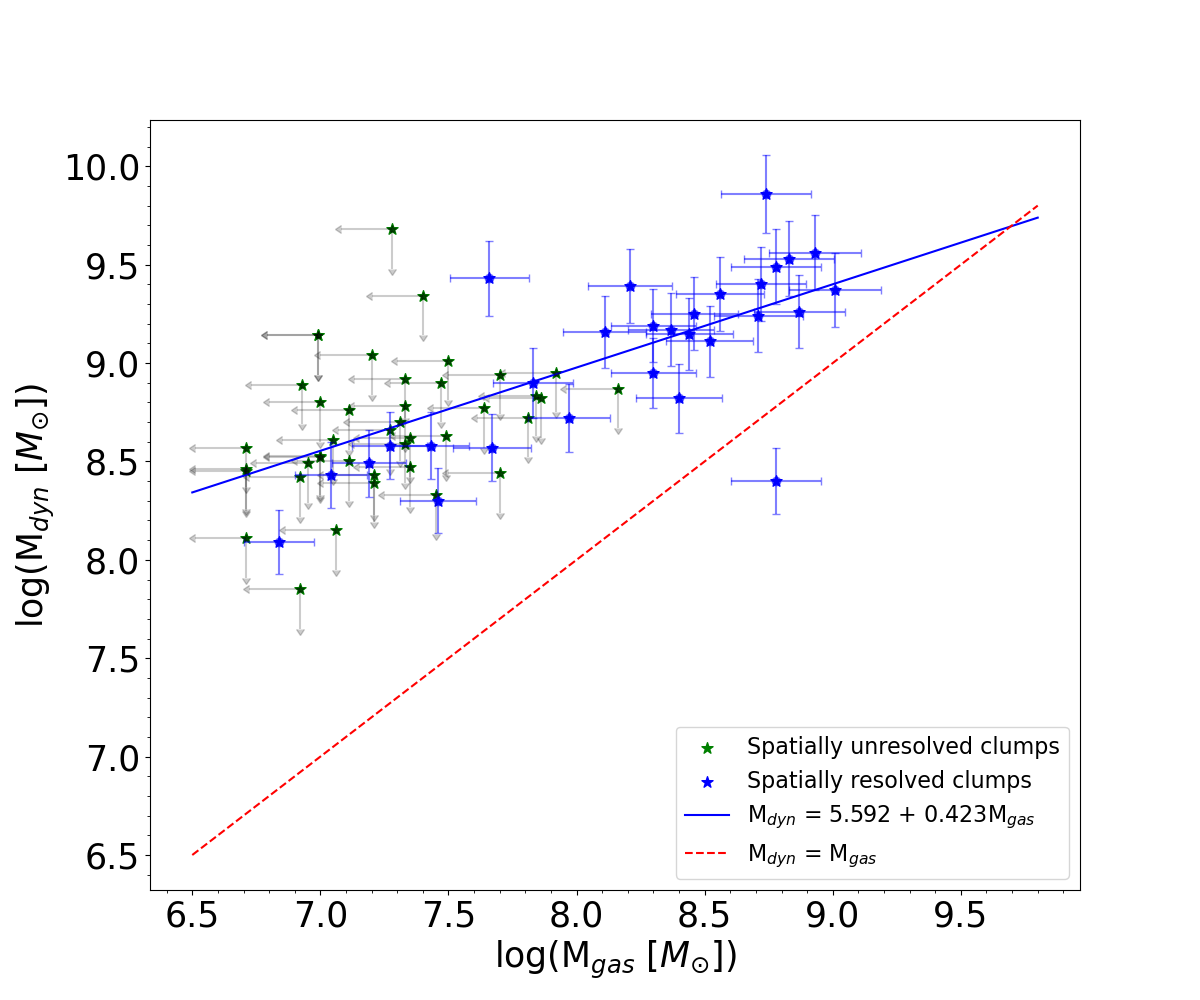}
\caption{ The relation between gas mass and dynamical mass identified in this work. The gas mass in the x axis was obtain using the Schmidt-Kennicutt relation. The dynamical gas was obtain using velocity dispersion and assuming a virialized objects. The blue slash dotted line is the linear regression and the green line represents equality between the masses. }
\label{fig:MgasxMdyn}

\end{figure}
\end{center}




\begin{center}
\begin{figure}
\includegraphics[width=\linewidth]{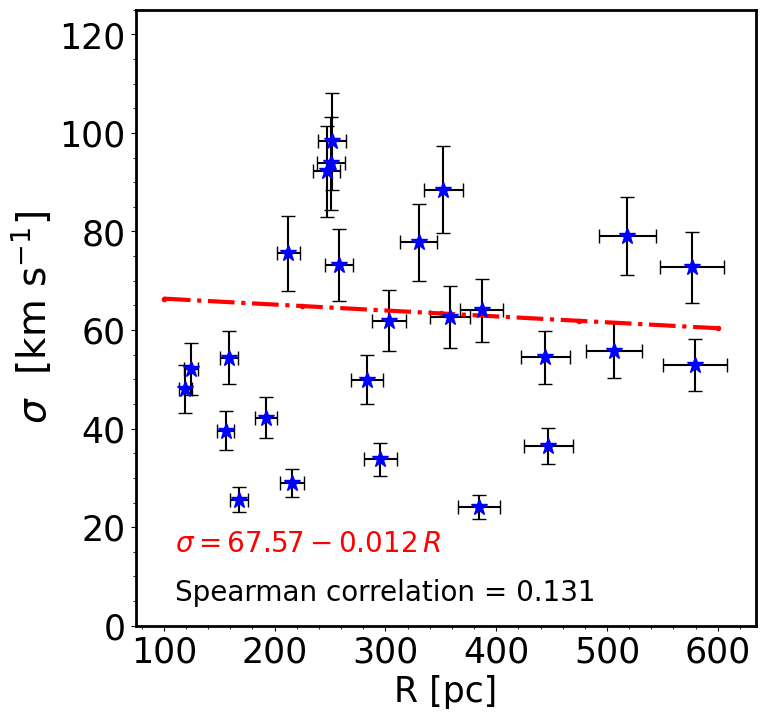}
\caption{ Relationship between the size of the clump and its dispersion of velocities($\sigma$) The red dotted line is the linear regression.}
\label{fig:Rxsigma}
\end{figure}
\end{center}
As can be seen in Figure \ref{fig:MgasxMdyn}, the dynamical mass of the selected clumps is directly correlated with the independently obtained gas mass, showing a Spearman rank correlation coefficient $\rho =$ 0.708, with a probability that both measurements are correlated greater than 99.9\%, as expected. Inferred dynamical masses are typically higher than gas masses, especially at the low-mass end of the relation, indicating that the contribution from star formation feedback might be proportionally higher for smaller clumps -- i.e., dynamical masses might be severely overestimated due to increased turbulency from stellar winds in these clumps. While stellar mass will certainly play a part, we argue that in LBAs -- especially within bluer \citep[as seen by their HST colors][]{Overzier2009}, highly star-forming regions -- the mass budget will be dominated by gas, as seen in \citet{Thiago2014}.


In figure \ref{fig:Rxsigma} we present the comparison between the size of the clump and its velocity dispersion. No correlation was found between these specific clump values. \cite{livermore2015} reached the same result for his high-redshift lensed clumps sample, suggesting that the velocity dispersion ($\sigma$) of the clumps studied by them may contain contributions from gravitational instabilities and star-forming feedback, since the clumps are not virialized. These hypotheses, in addition to feedback of supernovae that may be occurring, could explain the high scatter in the relation between the gas mass and the velocity dispersion.



\section{Results \& Discussion}

\subsection{Clump sizes}

Regardless of the model of clump formation, the uncertainty in the shape and size of star-forming clumps at cosmic noon has led to intense discussions over several years. \cite{Elmegreen2007} and \cite{Forster2011} found typical clump sizes above 1 kpc. Star-forming galaxies at redshifts between $0.1< z < 1$ show scales of 100 to several hundred parsecs \citep{Rozas2006, Fisher2017, lenkic2021} while works using cosmic noon galaxies magnified by gravitational lensing \citep{livermore2012, livermore2015, Cava2018, Dessauges2017a, Dessauges2018, Dessauges2019} show clumps down to sizes of a few tens of parsecs.

According to \cite{Fisher2017}, the discrepancy in size scale, with its potential overestimation, is attributed to resolution and observational sensitivity. However, \cite{cosens2018} suggests that it is invariant with respect to resolution. This author argues that, regardless of whether the images are observed with magnification from gravitational lenses or not, the relationship between H-alpha luminosity and the size scale of clumps remains constant.

Nevertheless, there is still intense debate about the actual size of clumps. One might argue that the main observational limitation is the instrumental angular resolution for a given redshift. For observations without the aid of adaptive optics (AO) and gravitational lensing, previous results by \cite{Elmegreen2007} and \cite{Forster2011} yield clump sizes greater than 1kpc. \cite{livermore2012,livermore2015} , \cite{Dessauges2017b,Dessauges2017a} and \cite{Cava2018}, however, determined clump dimensions of a few hundred parsecs or even smaller, at the same redshift, using observations of gravitationally lensed objects. 

\begin{figure*}
\begin{center}

\includegraphics[width=\linewidth]{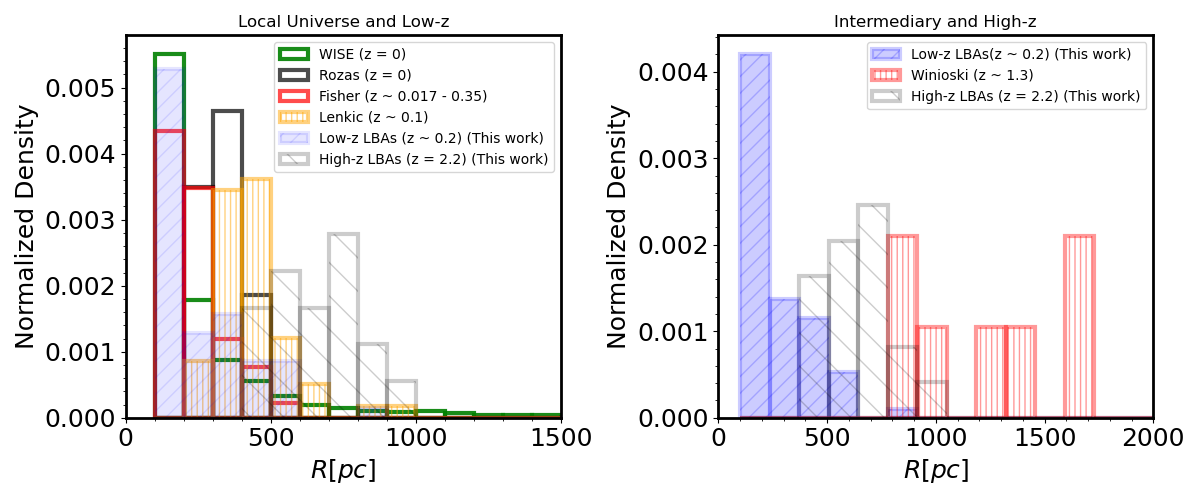}

\caption{Plot normalized by the number of data in each bin for the local HII regions from the WISE catalog \citep{Wright2010}, one sample selected from among the brightest and most isolated in a group of spiral galaxies \citep{Rozas2006}, clumps found by \citet{wisnioski2012} in redshift $\sim$ 1.3 from the WiggleZ Dark Energy Survey, a sample of giant extragalactic HII regions,  giant clumps in DYNAMO main sequence galaxies at z $\sim$ 0.017 - 0.35 \citep{Fisher2017}, giant clumps in DYNAMO galaxies at z $\sim$ 0.1 \citep{lenkic2021}, this work at z $\sim$ 0.2 and simulated high-z clumps at  z = 2.2.}
\label{fig:radius}
\end{center}
\end{figure*}

In order to determine the observational biases that might affect clump size measurements and contextualize our results, we compare the sizes of LBA clumps with local HII regions from the WISE catalog \citep{Wright2010}, one sample selected from among the brightest and most isolated in a group of spiral galaxies \citep{Rozas2006}, giant clumps in DYNAMO galaxies at z $\sim$ 0.1 \citep{lenkic2021}, a sample of giant extragalactic HII regions, giant clumps in DYNAMO main sequence galaxies at z $\sim$ 0.017 - 0.35 \citep{Fisher2017}, and clumps found by \cite{wisnioski2012} in z $\sim$ 1.3 from the WiggleZ Dark Energy Survey. In Figure \ref{fig:radius} we include both the direct results of our measurements at $z \sim 0.2$ and the simulated high-z observations at $z = 2.2$.

Upon examination of our sample at low-z, we conclude that the radii of the star-forming clumps are primarily in the range of a few hundred parsecs.  Regardless of the method used, be it through IFU observations or gravitational lensing, this reinforces the scenario in which clump radii exhibit typical scales smaller than 1 kpc. This finding alone suggests the presence of an inherent observational bias, since LBAs are similar to cosmic-noon galaxies in all other respects.

\begin{figure}
\begin{center}
\includegraphics[width=\linewidth]{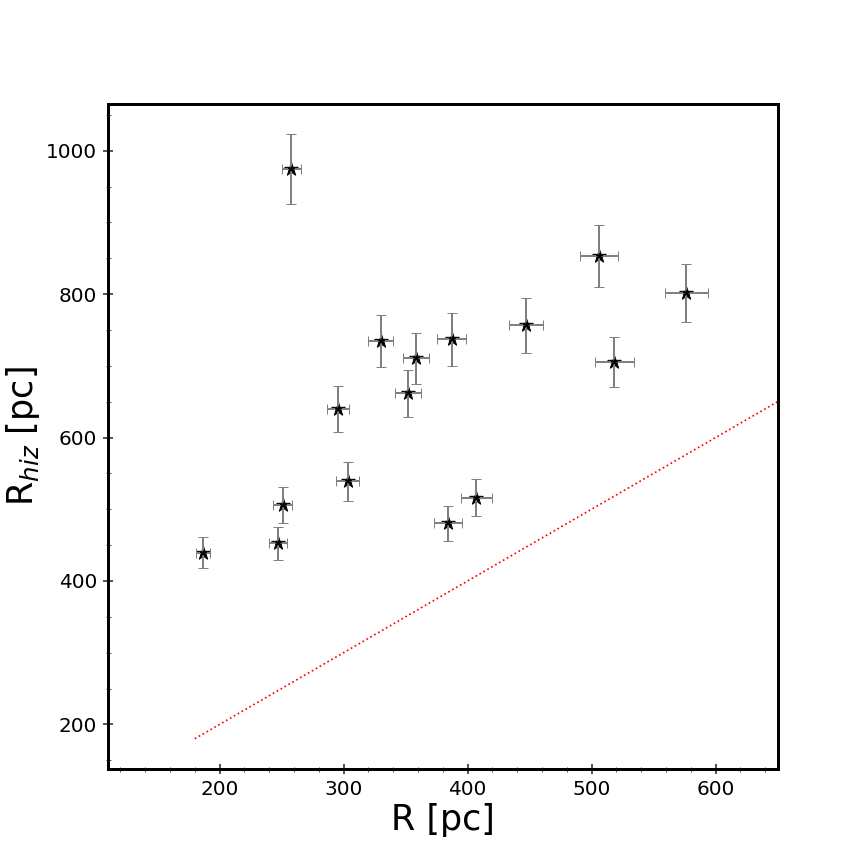}

\caption{ Plot showing clump sizes at z = 2.2 and z $\approx$ 0.2, for 16 corresponding clumps at the two distances. It was observed that all clumps in high-z showed an increase in their radii, even after the quadratic correction of the FWHM. The red dotted line represents equality between the two measures.} 
\label{fig:R_hizXR_local}
\end{center}
\end{figure}

We also included the measured clump sizes for the simulated high redshift observations, and the same effects found by \cite{Fisher2017} were observed. As seen in Figure \ref{fig:R_hizXR_local}, inferred radii are much larger at z=2.2, even after correcting for the larger FWHM. In addition, the number of clumps reduced from 84 in total to 17. This was probably due to the effect of clump clustering and also to the increase in the RMS of the image, which meant that some clumps were not identified by the criteria presented above. This agrees with results from \cite{Fisher2017}, who used H$\alpha$ images from the Hubble Space Telescope and detected the same effect of clump clustering, which systematically increases the apparent size of the clumps, causing them to be observed in high-z galaxies on kiloparsec scales and significantly affecting the measured properties of these objects.

To enable a direct comparison of the sizes and morphologies of our low- and high-redshift samples, we performed redshift simulations on the low‑z LBA sample. Corrections were applied to datacubes both for cosmological surface brightness dimming and for variations in physical resolution, to emulate observations of distant objects. We also assumed that all galaxies are observed in H$\alpha$ in the $H$ band at redshift $z=2.2$, away from any prominent sky lines. In this case, the flux is corrected according to total H$\alpha$ as measured with SDSS. Total line luminosities are typically 8 times higher for H$\alpha$, which would be expected for Case B recombination with little extinction. For more details, see \citet{Thiago2010}.

\begin{figure}
\begin{center}
\includegraphics[width=\linewidth]{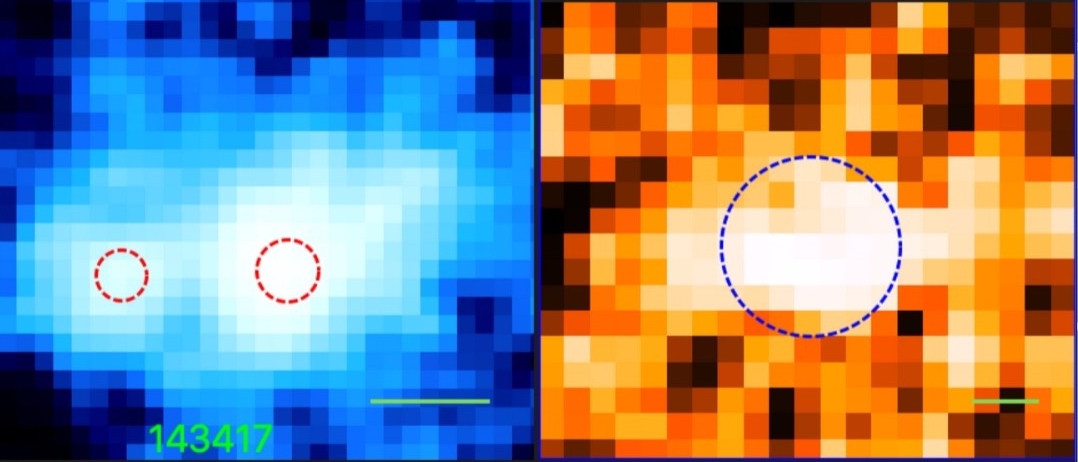}
\caption{ In the left image, UVLG 143407 is depicted in Low-z, revealing the presence of two discernible clumps marked by red dotted circles. In contrast, the right image displays a simulated High-z (z = 2.2) view of the galaxy, in which only one clump is observable, highlighted by a blue dotted circle. The green line in the lower left corner represents a scale of 1 kpc }
\label{fig:143407_local_hiz}
\end{center}
\end{figure}

Figure \ref{fig:143407_local_hiz} shows a comparison between the low-redshift and simulated high-redshift images of UVLG 143407, as an example of this phenomenon. At low redshift, two clumps, marked with red dotted circles, are clearly identified. At high redshift, however, the algorithm detects only one clump with a much larger radius (blue dotted circle), corroborating the idea that the increase in clump size might occur due to the clump clustering effect.

The values obtained for the average velocity shear of all clumps in high-z and low-z are, respectively, 16 km/s and 13 km/s. The average velocity dispersion for all clumps found in high-z and low-z are 64 km/s and 76 km/s, respectively. As we can see, despite the aforementioned differences in clump sizes, velocity shear and dispersion values remained similar when simulating observations at cosmic noon.



\subsection{Dynamically hot clumps}

Many clumpy galaxies display a dynamic structure similar to rotating disks with turbulent gas content \citep{ForsterSchreiber2009,Thiago2010,wisnioski2012,Fisher2017a}. The combination of these two components can result in the formation of large and highly active star-forming clumps because of dynamical instabilities. The self-gravitating disk nonaxisymmetric instability model is the prevailing theory for clump formation. According to this model, once a certain size is reached, the rotation stabilizes against fragmentation driven by gravity \citep{Fisher2017a}.

Numerical simulations of gas-rich disks with turbulence have shown the creation of clumps on a scale of approximately 1 kpc \citep{Bournaud2014}. Theoretical analyzes, both analytical and numerical, indicate that these clumps can migrate toward the center and merge to form a bulge within a timescale of approximately 0.5 to 1.0 billion years \citep{Bournaud2014}.

The success of the violent disk instability model in creating bulges through the inward migration of large clumps depends on the ability of the clumps to survive in the face of outflows generated by stellar winds, supernovae, and radiation pressure. It is widely believed that this star formation feedback is a crucial factor in the evolution of star-forming galaxies, even if secular bulge growth can also occur directly from the disk without clump migration \citep{Genzel2011}.

\begin{figure}
\begin{center}
\includegraphics[width=\linewidth]{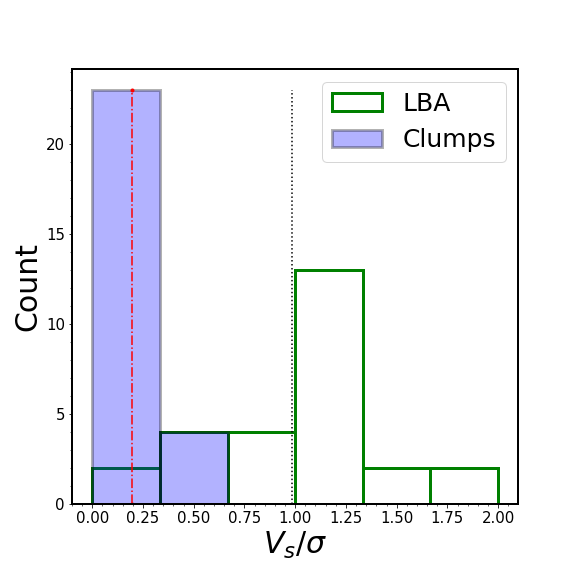}

\caption{Histograms showing the distribution of V$_s$/$\sigma$ relation of clumps of this work and the host LBAs \citep[as measured in][]{Thiago2010}. Green histograms show values for the entire galaxy, while shaded purple histograms indicate individual clumps. The red dotted line is the mean value of clumps (0.19) and the black dotted line is the mean value of V$_s$/$\sigma$ relation of host LBAs (0.98).} 
\label{fig:v/r}
\end{center}
\end{figure}

As an initial step, we examined the hypothesis that the clumps exist as separate structures within the disk of the host galaxy. We achieved this by analyzing the correlation between the shear velocity and velocity dispersion of both the clump dataset and the host galaxy. Figure \ref{fig:v/r} clearly illustrates a well-defined bimodal distribution, highlighting the turbulence within the star-forming clumps in our sample.

The reason behind the high turbulent velocities observed in star-forming galaxies and inside the clumps at high redshift remains an open question \citep{Fisher2017}. \cite{Lehnert2009} argues that the self-gravity model can be an important factor for the generation of turbulence but it is not enough to induce the observed velocity dispersion. This would also suggest that the injection of energy resulting from star formation is an important factor in generating such high values of velocity dispersion.

Using the obtained data, we determined the degree of instability of the clumps based on the theory of linear perturbation for asymmetric density waves, and estimated by the following equation:

\begin{equation}
Q = \frac{\sigma \kappa}{\Sigma G A}
\end{equation}
where $Q$ is the Toomre parameter $\sigma$, $\kappa$, $\Sigma$, $G$, and $A$ represent, respectively, the velocity dispersion, epicyclic frequency, gas surface density, gravitational constant, and a numerical factor that varies according to the object's composition, being 3.36 for stellar disks and 3.1416 for gas disks.

The epicyclic frequency is directly related to the circular frequency ($\kappa \propto \Omega$) for Keplerian orbits. The value of $\sqrt{3}$ was adopted for the proportionality constant as we used an approximation of a uniform disk for the clumps \citep{Dekel2020}. The circular frequency was obtained using the expression $\Omega = \frac{V_c}{R} = \left(\frac{GM}{R^3}\right)^{1/2}$, where $V_c$ is the circular velocity, as proposed by \cite{mandelker2016} and \cite{Oklopcic2017}.

We obtained values of the Toomre parameter (Q), for 31 resolved clumps and observed that 17 of them are dynamically unstable ($Q<1$) \citep{Toomre1964}. We compared the obtained values of the Toomre Q parameter between the clumps and their host galaxies and found that 29 out of the 31 analyzed clumps have this ratio below 1, suggesting that the clumps fragment more rapidly than the galaxies. Thus, considering purely dynamical processes, this faster fragmentation induced by strong random dispersion should lead to more effective star formation, as observed by \cite{Genzel2011} for subkiloparsec-scale star-forming clumps.

\subsection{Star formation rates and sizes}

We also examine the correlation between the star formation rate (SFR) and the sizes of the clumps. \cite{livermore2012,livermore2015} confirmed the correlation found by \cite{Kennicutt1998}, concluding that the surface brightness of nearby HII regions remains nearly constant as a function radius. This correlation remains at different redshifts, except in systems that show signs of merger and gravitational interaction, as discussed by \citet{bastian2006}.

\begin{figure*}
\begin{center}
\includegraphics[width=\linewidth]{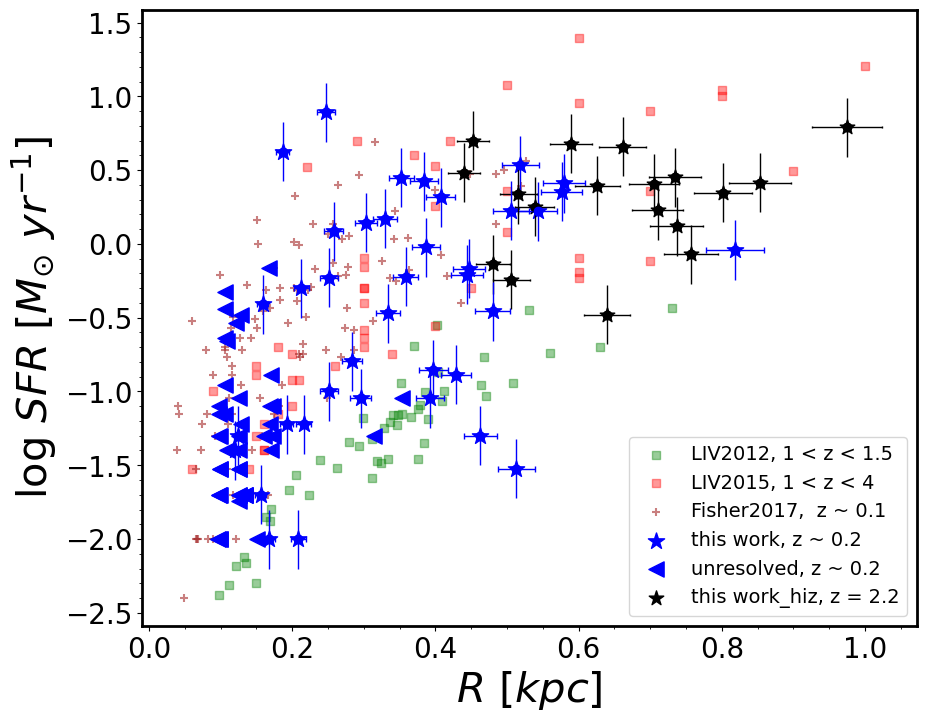}
\caption{Relation between SFR and R for different clump samples. The green squares represent the clumps located at redshifts between 1 - 1.5 \citep{livermore2012}, while the red filled squares represent clumps at redshifts between 1-4 \citep{livermore2015} and the crosses represent clumps at $z \sim 0.1$ \citep{Fisher2017}. The measurements for local clumps investigated in this work are represented by blue stars (spatially resolved clumps) and the blue triangles (spatially unresolved clumps). The black stars represent  the redshifted clumps at high-z. Our measurements at low redshift are in excellent agreement with other intermediate redshift clumps, while the simulated data at cosmic noon might be the result of the combination of individual, smaller clumps.}
\label{fig:sfrxr}
\end{center}
\end{figure*}

The size-SFR relationship has been well established in the local universe, particularly through the study of H II regions in nearby spiral and irregular galaxies, as detailed by \cite{Kennicutt1998}. In their comprehensive analysis, \cite{livermore2015} synthesized findings from multiple studies that included star-forming regions in typical main-sequence galaxies of the local universe and clumps in lensed galaxies with redshifts below 3. They observed that clumps with higher surface brightnesses are increasingly prevalent in galaxies at elevated redshifts. This pattern is attributed to the fact that such clumps emerge from fragmentation processes within galaxies that possess high gas fractions as redshift increases. Figure \ref{fig:sfrxr} shows a comparison between the data obtained from our low-z and high-z samples and the data from \cite{livermore2012,livermore2015} for clumps observed at 1 $<$ z $<$ 1.5 and 1 $<$ z $<$ 4, respectively, as well as the work on clumps in DYNAMO galaxies with z $\sim$ 0.1 by \citep{Fisher2017}. We notice that there is a correlation between the two quantities, with nearly constant surface densities of star formation for a given redshift, but higher radii and smaller densities per clump at higher redshifts.

We also compare our measured values with recent results for lensed galaxies with HST \citep{Mestric2022} and JWST \citep{Claeyssens2023,Claeyssens2025}. The excellent physical resolution achieved in those works due to strongly lensed galaxies in cluster fields allow for even more precise measurements than in this present work, but typical sizes of a few hundred pc and star formation rates of a few tenths of M$_\odot$ yr$^{-1}$ for galaxies beyond $z\gtrsim1$ are well matched with our data.

In summary, we argue that our LBA sample agrees well with these measurements, overlapping especially with samples at higher redshift (albeit with larger spread), reinforcing the argument for the cosmic noon analogy of LBAs. Our simulated observations at higher redshift still follow a similar relation at higher radii and star formation rate, supporting the idea that these are measurements of multiple clumps combined.



\section{Summary \& Conclusions}

We use data obtained with the OSIRIS Integral Field Unit (IFU) at the Keck telescope with the assistance of Laser Guide Star Adaptive Optics (LGS-AO) to analyze the properties of star-forming clumps in 18 Lyman Break Analogs at z $\sim 0.1-0.2$. Using the emission line in Pa$-{\alpha}$ of these objects we find and analyze 84 clumps with diameters on the scale of hundreds of parsecs. 
 Our main conclusions are summarized below.

(1) Clumps manifest as distinct structures within the disks of their host galaxies, with an average V/$\sigma$ ratio markedly dissimilar from the average value of this parameter exhibited by Lyman Break Analogs in our sample.

(2) Clumps exhibit radii on the order of a few hundred parsecs, larger than those of local HII regions but smaller than the clumps observed in galaxies at high redshift without the aid of gravitational lensing, which reveal clumps with radii in few kiloparsecs.  These results are consistent with numerous studies utilizing LGS-AO and its resolution power.

(3) The size of the clumps observed in the flux maps projected for z = 2.2 show higher values than those observed in the low-z clumps flux maps, even after correction for instrumental angular resolution. This is likely due to the clump clustering effect observed by the loss of resolution and the clump selection criteria.

(4) Based on the calculations of the Toomre Q parameter for each clump and LBAs, at least 50\% are dynamically unstable, and when compared to the host galaxies, 94\% of the clumps are even more unstable, with a potential of further fragmentation and the further increase of star formation efficiencies within each region.


(5) As can be seen in Figure \ref{fig:sfrxr}, the clump clustering effect, in addition to producing clumps with larger sizes than those observed in low-z, causes a significant increase in the observed SFR. This is probablly due to the combined fluxes of individual smaller clumps.

The implications of our findings underscore the importance of conducting further investigations into the properties of star-forming clumps across varying redshift regimes. By probing the spatial distribution, dynamics, and other physical characteristics of these clumps, we gain valuable insights into the processes governing galaxy formation and evolution. These insights are particularly crucial for developing a more comprehensive understanding of the mechanisms driving the evolution of galaxies across cosmic time. Therefore, the study of star-forming clumps across different redshift regimes represents a fundamental avenue for advancing our knowledge of the universe and its complex evolutionary history.

In addition to the conclusions presented above, this work corroborates the results obtained in previous works, such as \citep{Overzier2009, Overzier2010} and \citep{Thiago2010,Thiago2014}, which show that LBAs should be used to study the structures and components of star-forming galaxies at cosmic noon, as they are much closer, allowing for observations with a better special resolution. This spatial resolution limitation for cosmic noon galaxies probably could be compensated with the entry into operation of telescopes of the approximately 40 meter class, such as the upcoming Extremely Large Telescope (ELT).

\section*{Acknowledgments}

We would like to thank the anonymous referee for useful comments that significantly improved the scientific discussion in the paper. This study was financed in part by the Coordenação de Aperfeiçoamento de Pessoal de Nível Superior – Brasil (CAPES) – Finance Code 001. TSG would like to thank the support of CNPq (grant 314747/2020-6) and FAPERJ (grant E-26/201.309/2021). KMD also thanks the support of the Serrapilheira Institute (grant Serra-1709-17357) as well as that of the Brazilian National Research Council (CNPq grant 308584/2022-8) and of the Rio de Janeiro State Research Foundation (FAPERJ grant E-26/200.952/2022), Brazil. We would also like to thank those of Hawaiian ancestry for hospitably allowing telescope operations on the summit of Mauna Kea.

\bibliography{sample631}{}

\begin{thebibliography}{}
\expandafter\ifx\csname natexlab\endcsname\relax\def\natexlab#1{#1}\fi
\providecommand{\url}[1]{\href{#1}{#1}}
\providecommand{\dodoi}[1]{doi:~\href{http://doi.org/#1}{\nolinkurl{#1}}}
\providecommand{\doeprint}[1]{\href{http://ascl.net/#1}{\nolinkurl{http://ascl.net/#1}}}
\providecommand{\doarXiv}[1]{\href{https://arxiv.org/abs/#1}{\nolinkurl{https://arxiv.org/abs/#1}}}

\bibitem[{{Adamo} {et~al.}(2024){Adamo}, {Atek}, {Bagley}, {Ba{\~n}ados},
  {Barrow}, {Berg}, {Bezanson}, {Brada{\v{c}}}, {Brammer}, {Carnall},
  {Chisholm}, {Coe}, {Dayal}, {Eisenstein}, {Eldridge}, {Ferrara}, {Fujimoto},
  {de Graaff}, {Habouzit}, {Hutchison}, {Kartaltepe}, {Kassin}, {Kriek},
  {Labb{\'e}}, {Maiolino}, {Marques-Chaves}, {Maseda}, {Mason}, {Matthee},
  {McQuinn}, {Meynet}, {Naidu}, {Oesch}, {Pentericci},
  {P{\'e}rez-Gonz{\'a}lez}, {Rigby}, {Roberts-Borsani}, {Schaerer}, {Shapley},
  {Stark}, {Stiavelli}, {Strom}, {Vanzella}, {Wang}, {Wilkins}, {Williams},
  {Willott}, {Wylezalek}, \& {Nota}}]{Adamo2024}
{Adamo}, A., {Atek}, H., {Bagley}, M.~B., {et~al.} 2024, arXiv e-prints,
  arXiv:2405.21054, \dodoi{10.48550/arXiv.2405.21054}

\bibitem[{{Bastian} {et~al.}(2006){Bastian}, {Emsellem}, {Kissler-Patig}, \&
  {Maraston}}]{bastian2006}
{Bastian}, N., {Emsellem}, E., {Kissler-Patig}, M., \& {Maraston}, C. 2006,
  \aap, 445, 471, \dodoi{10.1051/0004-6361:20053793}

\bibitem[{{Basu-Zych} {et~al.}(2007){Basu-Zych}, {Schiminovich}, {Johnson},
  {Hoopes}, {Overzier}, {Treyer}, {Heckman}, {Barlow}, {Bianchi}, {Conrow},
  {Donas}, {Forster}, {Friedman}, {Lee}, {Madore}, {Martin}, {Milliard},
  {Morrissey}, {Neff}, {Rich}, {Salim}, {Seibert}, {Small}, {Szalay}, {Wyder},
  \& {Yi}}]{Basu-Zych2007}
{Basu-Zych}, A.~R., {Schiminovich}, D., {Johnson}, B.~D., {et~al.} 2007, \apjs,
  173, 457, \dodoi{10.1086/521146}

\bibitem[{{Basu-Zych} {et~al.}(2009){Basu-Zych}, {Gon{\c{c}}alves}, {Overzier},
  {Law}, {Schiminovich}, {Heckman}, {Martin}, {Wyder}, \&
  {O'Dowd}}]{Basu-Zych2009}
{Basu-Zych}, A.~R., {Gon{\c{c}}alves}, T.~S., {Overzier}, R., {et~al.} 2009,
  \apjl, 699, L118, \dodoi{10.1088/0004-637X/699/2/L118}

\bibitem[{{Basu-Zych} {et~al.}(2013){Basu-Zych}, {Lehmer}, {Hornschemeier},
  {Gon{\c{c}}alves}, {Fragos}, {Heckman}, {Overzier}, {Ptak}, \&
  {Schiminovich}}]{Basu-Zych2013}
{Basu-Zych}, A.~R., {Lehmer}, B.~D., {Hornschemeier}, A.~E., {et~al.} 2013,
  \apj, 774, 152, \dodoi{10.1088/0004-637X/774/2/152}

\bibitem[{{Behrendt} {et~al.}(2015){Behrendt}, {Burkert}, \&
  {Schartmann}}]{behrendt2015}
{Behrendt}, M., {Burkert}, A., \& {Schartmann}, M. 2015, \mnras, 448, 1007,
  \dodoi{10.1093/mnras/stv027}

\bibitem[{{Berry}(2015)}]{Berry2015}
{Berry}, D.~S. 2015, Astronomy and Computing, 10, 22,
  \dodoi{10.1016/j.ascom.2014.11.004}

\bibitem[{{Boada} {et~al.}(2015){Boada}, {Tilvi}, {Papovich}, {Quadri},
  {Hilton}, {Finkelstein}, {Guo}, {Bond}, {Conselice}, {Dekel}, {Ferguson},
  {Giavalisco}, {Grogin}, {Kocevski}, {Koekemoer}, \& {Koo}}]{Boada2015}
{Boada}, S., {Tilvi}, V., {Papovich}, C., {et~al.} 2015, \apj, 803, 104,
  \dodoi{10.1088/0004-637X/803/2/104}

\bibitem[{{Bournaud} {et~al.}(2008){Bournaud}, {Duc}, \&
  {Emsellem}}]{Bournaud2008}
{Bournaud}, F., {Duc}, P.~A., \& {Emsellem}, E. 2008, \mnras, 389, L8,
  \dodoi{10.1111/j.1745-3933.2008.00511.x}

\bibitem[{{Bournaud} {et~al.}(2007){Bournaud}, {Elmegreen}, \&
  {Elmegreen}}]{Bournaud2007}
{Bournaud}, F., {Elmegreen}, B.~G., \& {Elmegreen}, D.~M. 2007, \apj, 670, 237,
  \dodoi{10.1086/522077}

\bibitem[{{Bournaud} {et~al.}(2009){Bournaud}, {Elmegreen}, \&
  {Martig}}]{Bournaud&elmegreen2009}
{Bournaud}, F., {Elmegreen}, B.~G., \& {Martig}, M. 2009, \apjl, 707, L1,
  \dodoi{10.1088/0004-637X/707/1/L1}

\bibitem[{{Bournaud} {et~al.}(2014){Bournaud}, {Perret}, {Renaud}, {Dekel},
  {Elmegreen}, {Elmegreen}, {Teyssier}, {Amram}, {Daddi}, {Duc}, {Elbaz},
  {Epinat}, {Gabor}, {Juneau}, {Kraljic}, \& {Le Floch'}}]{Bournaud2014}
{Bournaud}, F., {Perret}, V., {Renaud}, F., {et~al.} 2014, \apj, 780, 57,
  \dodoi{10.1088/0004-637X/780/1/57}

\bibitem[{{Cava} {et~al.}(2018){Cava}, {Schaerer}, {Richard},
  {P{\'e}rez-Gonz{\'a}lez}, {Dessauges-Zavadsky}, {Mayer}, \&
  {Tamburello}}]{Cava2018}
{Cava}, A., {Schaerer}, D., {Richard}, J., {et~al.} 2018, Nature Astronomy, 2,
  76, \dodoi{10.1038/s41550-017-0295-x}

\bibitem[{{Claeyssens} {et~al.}(2025){Claeyssens}, {Adamo}, {Messa},
  {Dessauges-Zavadsky}, {Richard}, {Kramarenko}, {Matthee}, \&
  {Naidu}}]{Claeyssens2025}
{Claeyssens}, A., {Adamo}, A., {Messa}, M., {et~al.} 2025, \mnras,
  \dodoi{10.1093/mnras/staf058}

\bibitem[{{Claeyssens} {et~al.}(2023){Claeyssens}, {Adamo}, {Richard},
  {Mahler}, {Messa}, \& {Dessauges-Zavadsky}}]{Claeyssens2023}
{Claeyssens}, A., {Adamo}, A., {Richard}, J., {et~al.} 2023, \mnras, 520, 2180,
  \dodoi{10.1093/mnras/stac3791}

\bibitem[{{Conselice} {et~al.}(2022){Conselice}, {Mundy}, {Ferreira}, \&
  {Duncan}}]{Conselice2022}
{Conselice}, C.~J., {Mundy}, C.~J., {Ferreira}, L., \& {Duncan}, K. 2022, \apj,
  940, 168, \dodoi{10.3847/1538-4357/ac9b1a}

\bibitem[{{Contursi} {et~al.}(2017){Contursi}, {Baker}, {Berta}, {Magnelli},
  {Lutz}, {Fischer}, {Verma}, {Nielbock}, {Gr{\'a}cia Carpio}, {Veilleux},
  {Sturm}, {Davies}, {Genzel}, {Hailey-Dunsheath}, {Herrera-Camus}, {Janssen},
  {Poglitsch}, {Sternberg}, \& {Tacconi}}]{contursi2017}
{Contursi}, A., {Baker}, A.~J., {Berta}, S., {et~al.} 2017, \aap, 606, A86,
  \dodoi{10.1051/0004-6361/201730609}

\bibitem[{{Cosens} {et~al.}(2018){Cosens}, {Wright}, {Mieda}, {Murray},
  {Armus}, {Do}, {Larkin}, {Larson}, {Martinez}, {Walth}, \&
  {Vayner}}]{cosens2018}
{Cosens}, M., {Wright}, S.~A., {Mieda}, E., {et~al.} 2018, \apj, 869, 11,
  \dodoi{10.3847/1538-4357/aaeb8f}

\bibitem[{{Dekel} {et~al.}(2020){Dekel}, {Lapiner}, {Ginzburg}, {Freundlich},
  {Jiang}, {Finish}, {Kretschmer}, {Lin}, {Ceverino}, {Primack}, {Giavalisco},
  \& {Ji}}]{Dekel2020}
{Dekel}, A., {Lapiner}, S., {Ginzburg}, O., {et~al.} 2020, \mnras, 496, 5372,
  \dodoi{10.1093/mnras/staa1713}

\bibitem[{{Dessauges-Zavadsky} \& {Adamo}(2018)}]{Dessauges2018}
{Dessauges-Zavadsky}, M., \& {Adamo}, A. 2018, \mnras, 479, L118,
  \dodoi{10.1093/mnrasl/sly112}

\bibitem[{{Dessauges-Zavadsky}
  {et~al.}(2017{\natexlab{a}}){Dessauges-Zavadsky}, {Schaerer}, {Cava},
  {Mayer}, \& {Tamburello}}]{Dessauges2017a}
{Dessauges-Zavadsky}, M., {Schaerer}, D., {Cava}, A., {Mayer}, L., \&
  {Tamburello}, V. 2017{\natexlab{a}}, \apjl, 836, L22,
  \dodoi{10.3847/2041-8213/aa5d52}

\bibitem[{{Dessauges-Zavadsky}
  {et~al.}(2017{\natexlab{b}}){Dessauges-Zavadsky}, {Zamojski}, {Rujopakarn},
  {Richard}, {Sklias}, {Schaerer}, {Combes}, {Ebeling}, {Rawle}, {Egami},
  {Boone}, {Cl{\'e}ment}, {Kneib}, {Nyland}, \& {Walth}}]{Dessauges2017b}
{Dessauges-Zavadsky}, M., {Zamojski}, M., {Rujopakarn}, W., {et~al.}
  2017{\natexlab{b}}, \aap, 605, A81, \dodoi{10.1051/0004-6361/201628513}

\bibitem[{{Dessauges-Zavadsky} {et~al.}(2019){Dessauges-Zavadsky}, {Richard},
  {Combes}, {Schaerer}, {Rujopakarn}, {Mayer}, {Cava}, {Boone}, {Egami},
  {Kneib}, {P{\'e}rez-Gonz{\'a}lez}, {Pfenniger}, {Rawle}, {Teyssier}, \& {van
  der Werf}}]{Dessauges2019}
{Dessauges-Zavadsky}, M., {Richard}, J., {Combes}, F., {et~al.} 2019, Nature
  Astronomy, 3, 1115, \dodoi{10.1038/s41550-019-0874-0}

\bibitem[{{Elmegreen} {et~al.}(2009){Elmegreen}, {Elmegreen}, {Marcus},
  {Shahinyan}, {Yau}, \& {Petersen}}]{elmegreen2009a}
{Elmegreen}, D.~M., {Elmegreen}, B.~G., {Marcus}, M.~T., {et~al.} 2009, \apj,
  701, 306, \dodoi{10.1088/0004-637X/701/1/306}

\bibitem[{{Elmegreen} {et~al.}(2007){Elmegreen}, {Elmegreen}, {Ravindranath},
  \& {Coe}}]{Elmegreen2007}
{Elmegreen}, D.~M., {Elmegreen}, B.~G., {Ravindranath}, S., \& {Coe}, D.~A.
  2007, \apj, 658, 763, \dodoi{10.1086/511667}

\bibitem[{{Elmegreen} {et~al.}(2021){Elmegreen}, {Elmegreen}, {Whitmore},
  {Chandar}, {Calzetti}, {Lee}, {White}, {Cook}, {Ubeda}, {Mok}, \&
  {Linden}}]{elmegreen2021}
{Elmegreen}, D.~M., {Elmegreen}, B.~G., {Whitmore}, B.~C., {et~al.} 2021, \apj,
  908, 121, \dodoi{10.3847/1538-4357/abd541}

\bibitem[{{Erb} {et~al.}(2006){Erb}, {Steidel}, {Shapley}, {Pettini}, {Reddy},
  \& {Adelberger}}]{erb2006}
{Erb}, D.~K., {Steidel}, C.~C., {Shapley}, A.~E., {et~al.} 2006, \apj, 646,
  107, \dodoi{10.1086/504891}

\bibitem[{{Fisher} {et~al.}(2017{\natexlab{a}}){Fisher}, {Glazebrook},
  {Damjanov}, {Abraham}, {Obreschkow}, {Wisnioski}, {Bassett}, {Green}, \&
  {McGregor}}]{Fisher2017}
{Fisher}, D.~B., {Glazebrook}, K., {Damjanov}, I., {et~al.} 2017{\natexlab{a}},
  \mnras, 464, 491, \dodoi{10.1093/mnras/stw2281}

\bibitem[{{Fisher} {et~al.}(2017{\natexlab{b}}){Fisher}, {Glazebrook},
  {Damjanov}, {Abraham}, {Obreschkow}, {Wisnioski}, {Bassett}, {Green}, \&
  {McGregor}}]{Fisher2017a}
---. 2017{\natexlab{b}}, \mnras, 464, 491, \dodoi{10.1093/mnras/stw2281}

\bibitem[{{F{\"o}rster Schreiber} {et~al.}(2011){F{\"o}rster Schreiber},
  {Shapley}, {Erb}, {Genzel}, {Steidel}, {Bouch{\'e}}, {Cresci}, \&
  {Davies}}]{Forster2011}
{F{\"o}rster Schreiber}, N.~M., {Shapley}, A.~E., {Erb}, D.~K., {et~al.} 2011,
  \apj, 731, 65, \dodoi{10.1088/0004-637X/731/1/65}

\bibitem[{{F{\"o}rster Schreiber} {et~al.}(2009){F{\"o}rster Schreiber},
  {Genzel}, {Bouch{\'e}}, {Cresci}, {Davies}, {Buschkamp}, {Shapiro},
  {Tacconi}, {Hicks}, {Genel}, {Shapley}, {Erb}, {Steidel}, {Lutz},
  {Eisenhauer}, {Gillessen}, {Sternberg}, {Renzini}, {Cimatti}, {Daddi},
  {Kurk}, {Lilly}, {Kong}, {Lehnert}, {Nesvadba}, {Verma}, {McCracken},
  {Arimoto}, {Mignoli}, \& {Onodera}}]{ForsterSchreiber2009}
{F{\"o}rster Schreiber}, N.~M., {Genzel}, R., {Bouch{\'e}}, N., {et~al.} 2009,
  \apj, 706, 1364, \dodoi{10.1088/0004-637X/706/2/1364}

\bibitem[{{Freundlich} {et~al.}(2013){Freundlich}, {Combes}, {Tacconi},
  {Cooper}, {Genzel}, {Neri}, {Bolatto}, {Bournaud}, {Burkert}, {Cox}, {Davis},
  {F{\"o}rster Schreiber}, {Garcia-Burillo}, {Gracia-Carpio}, {Lutz}, {Naab},
  {Newman}, {Sternberg}, \& {Weiner}}]{Freundlich2013}
{Freundlich}, J., {Combes}, F., {Tacconi}, L.~J., {et~al.} 2013, \aap, 553,
  A130, \dodoi{10.1051/0004-6361/201220981}

\bibitem[{{Genzel} {et~al.}(2011){Genzel}, {Newman}, {Jones}, {F{\"o}rster
  Schreiber}, {Shapiro}, {Genel}, {Lilly}, {Renzini}, {Tacconi}, {Bouch{\'e}},
  {Burkert}, {Cresci}, {Buschkamp}, {Carollo}, {Ceverino}, {Davies}, {Dekel},
  {Eisenhauer}, {Hicks}, {Kurk}, {Lutz}, {Mancini}, {Naab}, {Peng},
  {Sternberg}, {Vergani}, \& {Zamorani}}]{Genzel2011}
{Genzel}, R., {Newman}, S., {Jones}, T., {et~al.} 2011, \apj, 733, 101,
  \dodoi{10.1088/0004-637X/733/2/101}

\bibitem[{{Gon{\c{c}}alves} {et~al.}(2014){Gon{\c{c}}alves}, {Basu-Zych},
  {Overzier}, {P{\'e}rez}, \& {Martin}}]{Thiago2014}
{Gon{\c{c}}alves}, T.~S., {Basu-Zych}, A., {Overzier}, R.~A., {P{\'e}rez}, L.,
  \& {Martin}, D.~C. 2014, \mnras, 442, 1429, \dodoi{10.1093/mnras/stu852}

\bibitem[{{Gon{\c{c}}alves} {et~al.}(2010){Gon{\c{c}}alves}, {Basu-Zych},
  {Overzier}, {Martin}, {Law}, {Schiminovich}, {Wyder}, {Mallery}, {Rich}, \&
  {Heckman}}]{Thiago2010}
{Gon{\c{c}}alves}, T.~S., {Basu-Zych}, A., {Overzier}, R., {et~al.} 2010, \apj,
  724, 1373, \dodoi{10.1088/0004-637X/724/2/1373}

\bibitem[{{Guo} {et~al.}(2015{\natexlab{a}}){Guo}, {Zheng}, {Wang}, \&
  {Fu}}]{Guo2015b}
{Guo}, K., {Zheng}, X.~Z., {Wang}, T., \& {Fu}, H. 2015{\natexlab{a}}, \apjl,
  808, L49, \dodoi{10.1088/2041-8205/808/2/L49}

\bibitem[{{Guo}(2012)}]{Guo2012}
{Guo}, Y. 2012, PhD thesis, University of Massachusetts Amherst, United States

\bibitem[{{Guo} {et~al.}(2015{\natexlab{b}}){Guo}, {Ferguson}, {Bell}, {Koo},
  {Conselice}, {Giavalisco}, {Kassin}, {Lu}, {Lucas}, {Mandelker}, {McIntosh},
  {Primack}, {Ravindranath}, {Barro}, {Ceverino}, {Dekel}, {Faber}, {Fang},
  {Koekemoer}, {Noeske}, {Rafelski}, \& {Straughn}}]{Guo2015a}
{Guo}, Y., {Ferguson}, H.~C., {Bell}, E.~F., {et~al.} 2015{\natexlab{b}}, \apj,
  800, 39, \dodoi{10.1088/0004-637X/800/1/39}

\bibitem[{{Heckman} {et~al.}(2005){Heckman}, {Hoopes}, {Seibert}, {Martin},
  {Salim}, {Rich}, {Kauffmann}, {Charlot}, {Barlow}, {Bianchi}, {Byun},
  {Donas}, {Forster}, {Friedman}, {Jelinsky}, {Lee}, {Madore}, {Malina},
  {Milliard}, {Morrissey}, {Neff}, {Schiminovich}, {Siegmund}, {Small},
  {Szalay}, {Welsh}, \& {Wyder}}]{Heckman2005a}
{Heckman}, T.~M., {Hoopes}, C.~G., {Seibert}, M., {et~al.} 2005, \apjl, 619,
  L35, \dodoi{10.1086/425979}

\bibitem[{{Hoopes} {et~al.}(2007){Hoopes}, {Heckman}, {Salim}, {Seibert},
  {Tremonti}, {Schiminovich}, {Rich}, {Martin}, {Charlot}, {Kauffmann},
  {Forster}, {Friedman}, {Morrissey}, {Neff}, {Small}, {Wyder}, {Bianchi},
  {Donas}, {Lee}, {Madore}, {Milliard}, {Szalay}, {Welsh}, \&
  {Yi}}]{Hoopes2007}
{Hoopes}, C.~G., {Heckman}, T.~M., {Salim}, S., {et~al.} 2007, \apjs, 173, 441,
  \dodoi{10.1086/516644}

\bibitem[{{Hopkins} {et~al.}(2010){Hopkins}, {Bundy}, {Croton}, {Hernquist},
  {Keres}, {Khochfar}, {Stewart}, {Wetzel}, \& {Younger}}]{Hopkins2010}
{Hopkins}, P.~F., {Bundy}, K., {Croton}, D., {et~al.} 2010, \apj, 715, 202,
  \dodoi{10.1088/0004-637X/715/1/202}

\bibitem[{{Hung} {et~al.}(2016){Hung}, {Hayward}, {Smith}, {Ashby}, {Lanz},
  {Mart{\'\i}nez-Galarza}, {Sanders}, \& {Zezas}}]{Hunh2016}
{Hung}, C.-L., {Hayward}, C.~C., {Smith}, H.~A., {et~al.} 2016, \apj, 816, 99,
  \dodoi{10.3847/0004-637X/816/2/99}

\bibitem[{{Immeli} {et~al.}(2004){Immeli}, {Samland}, {Gerhard}, \&
  {Westera}}]{Immeli2004}
{Immeli}, A., {Samland}, M., {Gerhard}, O., \& {Westera}, P. 2004, \aap, 413,
  547, \dodoi{10.1051/0004-6361:20034282}

\bibitem[{{Kennicutt}(1998)}]{Kennicutt1998}
{Kennicutt}, Robert~C., J. 1998, \apj, 498, 541, \dodoi{10.1086/305588}

\bibitem[{{Kewley} {et~al.}(2005){Kewley}, {Jansen}, \& {Geller}}]{Kewley2005}
{Kewley}, L.~J., {Jansen}, R.~A., \& {Geller}, M.~J. 2005, \pasp, 117, 227,
  \dodoi{10.1086/428303}

\bibitem[{{Larkin} {et~al.}(2006){Larkin}, {Barczys}, {Krabbe}, {Adkins},
  {Aliado}, {Amico}, {Brims}, {Campbell}, {Canfield}, {Gasaway}, {Honey},
  {Iserlohe}, {Johnson}, {Kress}, {LaFreniere}, {Lyke}, {Magnone}, {Magnone},
  {McElwain}, {Moon}, {Quirrenbach}, {Skulason}, {Song}, {Spencer}, {Weiss}, \&
  {Wright}}]{Larkin2006}
{Larkin}, J., {Barczys}, M., {Krabbe}, A., {et~al.} 2006, in Society of
  Photo-Optical Instrumentation Engineers (SPIE) Conference Series, Vol. 6269,
  Society of Photo-Optical Instrumentation Engineers (SPIE) Conference Series,
  ed. I.~S. {McLean} \& M.~{Iye}, 62691A, \dodoi{10.1117/12.672061}

\bibitem[{{Law} {et~al.}(2006){Law}, {Steidel}, \& {Erb}}]{Law2006}
{Law}, D.~R., {Steidel}, C.~C., \& {Erb}, D.~K. 2006, \aj, 131, 70,
  \dodoi{10.1086/498683}

\bibitem[{{Law} {et~al.}(2007){Law}, {Steidel}, {Erb}, {Larkin}, {Pettini},
  {Shapley}, \& {Wright}}]{Law2007}
{Law}, D.~R., {Steidel}, C.~C., {Erb}, D.~K., {et~al.} 2007, \apj, 669, 929,
  \dodoi{10.1086/521786}

\bibitem[{{Law} {et~al.}(2009){Law}, {Steidel}, {Erb}, {Larkin}, {Pettini},
  {Shapley}, \& {Wright}}]{Law2009}
---. 2009, \apj, 697, 2057, \dodoi{10.1088/0004-637X/697/2/2057}

\bibitem[{{Lehnert} {et~al.}(2009){Lehnert}, {Nesvadba}, {Le Tiran}, {Di
  Matteo}, {van Driel}, {Douglas}, {Chemin}, \& {Bournaud}}]{Lehnert2009}
{Lehnert}, M.~D., {Nesvadba}, N.~P.~H., {Le Tiran}, L., {et~al.} 2009, \apj,
  699, 1660, \dodoi{10.1088/0004-637X/699/2/1660}

\bibitem[{{Lenki{\'c}} {et~al.}(2021){Lenki{\'c}}, {Bolatto}, {Fisher},
  {Glazebrook}, {Obreschkow}, {Abraham}, \& {Ambachew}}]{lenkic2021}
{Lenki{\'c}}, L., {Bolatto}, A.~D., {Fisher}, D.~B., {et~al.} 2021, \mnras,
  506, 3916, \dodoi{10.1093/mnras/stab1954}

\bibitem[{{Li} {et~al.}(2020){Li}, {Wang}, {Wu}, {Ma}, \& {Lin}}]{Li2020}
{Li}, C., {Wang}, H.-C., {Wu}, Y.-W., {Ma}, Y.-H., \& {Lin}, L.-H. 2020,
  Research in Astronomy and Astrophysics, 20, 031,
  \dodoi{10.1088/1674-4527/20/3/31}

\bibitem[{{Liu} {et~al.}(2020){Liu}, {Evans}, {Kim}, {Goldsmith}, {Liu},
  {Zhang}, {Tatematsu}, {Wang}, {Juvela}, {Bronfman}, {Cunningham}, {Garay},
  {Hirota}, {Lee}, {Kang}, {Li}, {Li}, {Mardones}, {Qin}, {Ristorcelli}, {Tej},
  {Toth}, {Wu}, {Wu}, {Yi}, {Yun}, {Liu}, {Peng}, {Li}, {Li}, {Lee}, {Shen},
  {Baug}, {Wang}, {Zhang}, {Issac}, {Zhu}, {Luo}, {Soam}, {Liu}, {Xu}, {Wang},
  {Zhang}, {Ren}, \& {Zhang}}]{Liu2020}
{Liu}, T., {Evans}, N.~J., {Kim}, K.-T., {et~al.} 2020, \mnras, 496, 2790,
  \dodoi{10.1093/mnras/staa1577}

\bibitem[{{Livermore} {et~al.}(2012){Livermore}, {Jones}, {Richard}, {Bower},
  {Ellis}, {Swinbank}, {Rigby}, {Smail}, {Arribas}, {Rodriguez Zaurin},
  {Colina}, {Ebeling}, \& {Crain}}]{livermore2012}
{Livermore}, R.~C., {Jones}, T., {Richard}, J., {et~al.} 2012, \mnras, 427,
  688, \dodoi{10.1111/j.1365-2966.2012.21900.x}

\bibitem[{{Livermore} {et~al.}(2015){Livermore}, {Jones}, {Richard}, {Bower},
  {Swinbank}, {Yuan}, {Edge}, {Ellis}, {Kewley}, {Smail}, {Coppin}, \&
  {Ebeling}}]{livermore2015}
{Livermore}, R.~C., {Jones}, T.~A., {Richard}, J., {et~al.} 2015, \mnras, 450,
  1812, \dodoi{10.1093/mnras/stv686}

\bibitem[{{Loaiza-Agudelo} {et~al.}(2020){Loaiza-Agudelo}, {Overzier}, \&
  {Heckman}}]{Loaiza-Agudelo2020}
{Loaiza-Agudelo}, M., {Overzier}, R.~A., \& {Heckman}, T.~M. 2020, \apj, 891,
  19, \dodoi{10.3847/1538-4357/ab6f6b}

\bibitem[{{Mandelker} {et~al.}(2016){Mandelker}, {Padnos}, {Dekel}, {Birnboim},
  {Burkert}, {Krumholz}, \& {Steinberg}}]{mandelker2016}
{Mandelker}, N., {Padnos}, D., {Dekel}, A., {et~al.} 2016, \mnras, 463, 3921,
  \dodoi{10.1093/mnras/stw2267}

\bibitem[{{Martin} {et~al.}(2005){Martin}, {Fanson}, {Schiminovich},
  {Morrissey}, {Friedman}, {Barlow}, {Conrow}, {Grange}, {Jelinsky},
  {Milliard}, {Siegmund}, {Bianchi}, {Byun}, {Donas}, {Forster}, {Heckman},
  {Lee}, {Madore}, {Malina}, {Neff}, {Rich}, {Small}, {Surber}, {Szalay},
  {Welsh}, \& {Wyder}}]{Martin2005}
{Martin}, D.~C., {Fanson}, J., {Schiminovich}, D., {et~al.} 2005, \apjl, 619,
  L1, \dodoi{10.1086/426387}

\bibitem[{{Me{\v{s}}tri{\'c}} {et~al.}(2022){Me{\v{s}}tri{\'c}}, {Vanzella},
  {Zanella}, {Castellano}, {Calura}, {Rosati}, {Bergamini}, {Mercurio},
  {Meneghetti}, {Grillo}, {Caminha}, {Nonino}, {Merlin}, {Cupani}, \&
  {Sani}}]{Mestric2022}
{Me{\v{s}}tri{\'c}}, U., {Vanzella}, E., {Zanella}, A., {et~al.} 2022, \mnras,
  516, 3532, \dodoi{10.1093/mnras/stac2309}

\bibitem[{{Mowla} {et~al.}(2022){Mowla}, {Iyer}, {Desprez},
  {Estrada-Carpenter}, {Martis}, {Noirot}, {Sarrouh}, {Strait}, {Asada},
  {Abraham}, {Brammer}, {Sawicki}, {Willott}, {Bradac}, {Doyon}, {Muzzin},
  {Pacifici}, {Ravindranath}, \& {Zabl}}]{Mowla2022}
{Mowla}, L., {Iyer}, K.~G., {Desprez}, G., {et~al.} 2022, \apjl, 937, L35,
  \dodoi{10.3847/2041-8213/ac90ca}

\bibitem[{{Mowla} {et~al.}(2024){Mowla}, {Iyer}, {Asada}, {Desprez}, {Tan},
  {Martis}, {Sarrouh}, {Strait}, {Abraham}, {Brada{\v{c}}}, {Brammer},
  {Muzzin}, {Pacifici}, {Ravindranath}, {Sawicki}, {Willott},
  {Estrada-Carpenter}, {Jahan}, {Noirot}, {Matharu}, {Rihtar{\v{s}}i{\v{c}}},
  \& {Zabl}}]{Mowla2024}
{Mowla}, L., {Iyer}, K., {Asada}, Y., {et~al.} 2024, arXiv e-prints,
  arXiv:2402.08696, \dodoi{10.48550/arXiv.2402.08696}

\bibitem[{{Nagy} {et~al.}(2023){Nagy}, {Dessauges-Zavadsky}, {Messa},
  {Richard}, {Sun}, {Combes}, \& {Eyholzer}}]{Nagy2023}
{Nagy}, D., {Dessauges-Zavadsky}, M., {Messa}, M., {et~al.} 2023, \aap, 678,
  A183, \dodoi{10.1051/0004-6361/202346951}

\bibitem[{{Noguchi}(1999)}]{Noguchi1999}
{Noguchi}, M. 1999, \apj, 514, 77, \dodoi{10.1086/306932}

\bibitem[{{Oklop{\v{c}}i{\'c}} {et~al.}(2017){Oklop{\v{c}}i{\'c}}, {Hopkins},
  {Feldmann}, {Kere{\v{s}}}, {Faucher-Gigu{\`e}re}, \& {Murray}}]{Oklopcic2017}
{Oklop{\v{c}}i{\'c}}, A., {Hopkins}, P.~F., {Feldmann}, R., {et~al.} 2017,
  \mnras, 465, 952, \dodoi{10.1093/mnras/stw2754}

\bibitem[{{Overzier} {et~al.}(2010){Overzier}, {Heckman}, {Schiminovich},
  {Basu-Zych}, {Gon{\c{c}}alves}, {Martin}, \& {Rich}}]{Overzier2010}
{Overzier}, R.~A., {Heckman}, T.~M., {Schiminovich}, D., {et~al.} 2010, \apj,
  710, 979, \dodoi{10.1088/0004-637X/710/2/979}

\bibitem[{{Overzier} {et~al.}(2009){Overzier}, {Heckman}, {Tremonti}, {Armus},
  {Basu-Zych}, {Gon{\c{c}}alves}, {Rich}, {Martin}, {Ptak}, {Schiminovich},
  {Ford}, {Madore}, \& {Seibert}}]{Overzier2009}
{Overzier}, R.~A., {Heckman}, T.~M., {Tremonti}, C., {et~al.} 2009, \apj, 706,
  203, \dodoi{10.1088/0004-637X/706/1/203}

\bibitem[{{Overzier} {et~al.}(2011){Overzier}, {Heckman}, {Wang}, {Armus},
  {Buat}, {Howell}, {Meurer}, {Seibert}, {Siana}, {Basu-Zych}, {Charlot},
  {Gon{\c{c}}alves}, {Martin}, {Neill}, {Rich}, {Salim}, \&
  {Schiminovich}}]{Overzier2011}
{Overzier}, R.~A., {Heckman}, T.~M., {Wang}, J., {et~al.} 2011, \apjl, 726, L7,
  \dodoi{10.1088/2041-8205/726/1/L7}

\bibitem[{{Peng} {et~al.}(2010){Peng}, {Ho}, {Impey}, \& {Rix}}]{Peng2010}
{Peng}, C.~Y., {Ho}, L.~C., {Impey}, C.~D., \& {Rix}, H.-W. 2010, \aj, 139,
  2097, \dodoi{10.1088/0004-6256/139/6/2097}

\bibitem[{{Posses} {et~al.}(2024){Posses}, {Aravena}, {Gonz{\'a}lez-L{\'o}pez},
  {F{\"o}rster Schreiber}, {Liu}, {Lee}, {Solimano}, {D{\'\i}az-Santos},
  {Assef}, {Barcos-Mu{\~n}oz}, {Bovino}, {Bowler}, {Calistro Rivera}, {da
  Cunha}, {Davies}, {Killi}, {De Looze}, {Ferrara}, {Fisher}, {Herrera-Camus},
  {Ikeda}, {Lambert}, {Li}, {Lutz}, {Mitsuhashi}, {Palla}, {Rela{\~n}o},
  {Spilker}, {Naab}, {Tadaki}, {Telikova}, {{\"U}bler}, {van der Giessen}, \&
  {Villanueva}}]{Posses2024}
{Posses}, A., {Aravena}, M., {Gonz{\'a}lez-L{\'o}pez}, J., {et~al.} 2024, arXiv
  e-prints, arXiv:2403.03379, \dodoi{10.48550/arXiv.2403.03379}

\bibitem[{{Rigby} {et~al.}(2016){Rigby}, {Moore}, {Plume}, {Eden}, {Urquhart},
  {Thompson}, {Mottram}, {Brunt}, {Butner}, {Dempsey}, {Gibson}, {Hatchell},
  {Jenness}, {Kuno}, {Longmore}, {Morgan}, {Polychroni}, {Thomas}, {White}, \&
  {Zhu}}]{Rigby2016}
{Rigby}, A.~J., {Moore}, T.~J.~T., {Plume}, R., {et~al.} 2016, \mnras, 456,
  2885, \dodoi{10.1093/mnras/stv2808}

\bibitem[{{Rozas} {et~al.}(2006){Rozas}, {Richer}, {L{\'o}pez}, {Rela{\~n}o},
  \& {Beckman}}]{Rozas2006}
{Rozas}, M., {Richer}, M.~G., {L{\'o}pez}, J.~A., {Rela{\~n}o}, M., \&
  {Beckman}, J.~E. 2006, \aap, 455, 539, \dodoi{10.1051/0004-6361:20054388}

\bibitem[{{Somerville} {et~al.}(2001){Somerville}, {Primack}, \&
  {Faber}}]{Somerville2001}
{Somerville}, R.~S., {Primack}, J.~R., \& {Faber}, S.~M. 2001, \mnras, 320,
  504, \dodoi{10.1046/j.1365-8711.2001.03975.x}

\bibitem[{{Soto} {et~al.}(2017){Soto}, {de Mello}, {Rafelski}, {Gardner},
  {Teplitz}, {Koekemoer}, {Ravindranath}, {Grogin}, {Scarlata}, {Kurczynski},
  \& {Gawiser}}]{soto2017}
{Soto}, E., {de Mello}, D.~F., {Rafelski}, M., {et~al.} 2017, \apj, 837, 6,
  \dodoi{10.3847/1538-4357/aa5da3}

\bibitem[{{Spilker} {et~al.}(2022){Spilker}, {Hayward}, {Marrone}, {Aravena},
  {B{\'e}thermin}, {Burgoyne}, {Chapman}, {Greve}, {Gururajan}, {Hezaveh},
  {Hill}, {Litke}, {Lovell}, {Malkan}, {Murphy}, {Narayanan}, {Phadke},
  {Reuter}, {Stark}, {Sulzenauer}, {Vieira}, {Vizgan}, \&
  {Wei{\ss}}}]{Spilker2022}
{Spilker}, J.~S., {Hayward}, C.~C., {Marrone}, D.~P., {et~al.} 2022, \apjl,
  929, L3, \dodoi{10.3847/2041-8213/ac61e6}

\bibitem[{{Tacconi} {et~al.}(2010){Tacconi}, {Genzel}, {Neri}, {Cox}, {Cooper},
  {Shapiro}, {Bolatto}, {Bouch{\'e}}, {Bournaud}, {Burkert}, {Combes},
  {Comerford}, {Davis}, {F{\"o}rster Schreiber}, {Garcia-Burillo},
  {Gracia-Carpio}, {Lutz}, {Naab}, {Omont}, {Shapley}, {Sternberg}, \&
  {Weiner}}]{TAcconi2010}
{Tacconi}, L.~J., {Genzel}, R., {Neri}, R., {et~al.} 2010, \nat, 463, 781,
  \dodoi{10.1038/nature08773}

\bibitem[{{Toomre}(1964)}]{Toomre1964}
{Toomre}, A. 1964, \apj, 139, 1217, \dodoi{10.1086/147861}

\bibitem[{{Vanzella} {et~al.}(2023){Vanzella}, {Claeyssens}, {Welch}, {Adamo},
  {Coe}, {Diego}, {Mahler}, {Khullar}, {Kokorev}, {Oguri}, {Ravindranath},
  {Furtak}, {Hsiao}, {Abdurro'uf}, {Mandelker}, {Brammer}, {Bradley},
  {Brada{\v{c}}}, {Conselice}, {Dayal}, {Nonino}, {Andrade-Santos},
  {Windhorst}, {Pirzkal}, {Sharon}, {de Mink}, {Fujimoto}, {Zitrin},
  {Eldridge}, \& {Norman}}]{Vanzella2023}
{Vanzella}, E., {Claeyssens}, A., {Welch}, B., {et~al.} 2023, \apj, 945, 53,
  \dodoi{10.3847/1538-4357/acb59a}

\bibitem[{{Whitaker} {et~al.}(2012){Whitaker}, {van Dokkum}, {Brammer}, \&
  {Franx}}]{Whitaker2012}
{Whitaker}, K.~E., {van Dokkum}, P.~G., {Brammer}, G., \& {Franx}, M. 2012,
  \apjl, 754, L29, \dodoi{10.1088/2041-8205/754/2/L29}

\bibitem[{{Wisnioski} {et~al.}(2012){Wisnioski}, {Glazebrook}, {Blake},
  {Poole}, {Green}, {Wyder}, \& {Martin}}]{wisnioski2012}
{Wisnioski}, E., {Glazebrook}, K., {Blake}, C., {et~al.} 2012, \mnras, 422,
  3339, \dodoi{10.1111/j.1365-2966.2012.20850.x}

\bibitem[{{Wright} {et~al.}(2010){Wright}, {Eisenhardt}, {Mainzer}, {Ressler},
  {Cutri}, {Jarrett}, {Kirkpatrick}, {Padgett}, {McMillan}, {Skrutskie},
  {Stanford}, {Cohen}, {Walker}, {Mather}, {Leisawitz}, {Gautier}, {McLean},
  {Benford}, {Lonsdale}, {Blain}, {Mendez}, {Irace}, {Duval}, {Liu}, {Royer},
  {Heinrichsen}, {Howard}, {Shannon}, {Kendall}, {Walsh}, {Larsen}, {Cardon},
  {Schick}, {Schwalm}, {Abid}, {Fabinsky}, {Naes}, \& {Tsai}}]{Wright2010}
{Wright}, E.~L., {Eisenhardt}, P. R.~M., {Mainzer}, A.~K., {et~al.} 2010, \aj,
  140, 1868, \dodoi{10.1088/0004-6256/140/6/1868}

\bibitem[{{Wright} {et~al.}(2009){Wright}, {Larkin}, {Law}, {Steidel},
  {Shapley}, \& {Erb}}]{Wright2009}
{Wright}, S.~A., {Larkin}, J.~E., {Law}, D.~R., {et~al.} 2009, \apj, 699, 421,
  \dodoi{10.1088/0004-637X/699/1/421}

\bibitem[{{York} {et~al.}(2000){York}, {Adelman}, {Anderson}, {Anderson},
  {Annis}, {Bahcall}, {Bakken}, {Barkhouser}, {Bastian}, {Berman}, {Boroski},
  {Bracker}, {Briegel}, {Briggs}, {Brinkmann}, {Brunner}, {Burles}, {Carey},
  {Carr}, {Castander}, {Chen}, {Colestock}, {Connolly}, {Crocker}, {Csabai},
  {Czarapata}, {Davis}, {Doi}, {Dombeck}, {Eisenstein}, {Ellman}, {Elms},
  {Evans}, {Fan}, {Federwitz}, {Fiscelli}, {Friedman}, {Frieman}, {Fukugita},
  {Gillespie}, {Gunn}, {Gurbani}, {de Haas}, {Haldeman}, {Harris}, {Hayes},
  {Heckman}, {Hennessy}, {Hindsley}, {Holm}, {Holmgren}, {Huang}, {Hull},
  {Husby}, {Ichikawa}, {Ichikawa}, {Ivezi{\'c}}, {Kent}, {Kim}, {Kinney},
  {Klaene}, {Kleinman}, {Kleinman}, {Knapp}, {Korienek}, {Kron}, {Kunszt},
  {Lamb}, {Lee}, {Leger}, {Limmongkol}, {Lindenmeyer}, {Long}, {Loomis},
  {Loveday}, {Lucinio}, {Lupton}, {MacKinnon}, {Mannery}, {Mantsch}, {Margon},
  {McGehee}, {McKay}, {Meiksin}, {Merelli}, {Monet}, {Munn}, {Narayanan},
  {Nash}, {Neilsen}, {Neswold}, {Newberg}, {Nichol}, {Nicinski}, {Nonino},
  {Okada}, {Okamura}, {Ostriker}, {Owen}, {Pauls}, {Peoples}, {Peterson},
  {Petravick}, {Pier}, {Pope}, {Pordes}, {Prosapio}, {Rechenmacher}, {Quinn},
  {Richards}, {Richmond}, {Rivetta}, {Rockosi}, {Ruthmansdorfer}, {Sandford},
  {Schlegel}, {Schneider}, {Sekiguchi}, {Sergey}, {Shimasaku}, {Siegmund},
  {Smee}, {Smith}, {Snedden}, {Stone}, {Stoughton}, {Strauss}, {Stubbs},
  {SubbaRao}, {Szalay}, {Szapudi}, {Szokoly}, {Thakar}, {Tremonti}, {Tucker},
  {Uomoto}, {Vanden Berk}, {Vogeley}, {Waddell}, {Wang}, {Watanabe},
  {Weinberg}, {Yanny}, {Yasuda}, \& {SDSS Collaboration}}]{York2000}
{York}, D.~G., {Adelman}, J., {Anderson}, John~E., J., {et~al.} 2000, \aj, 120,
  1579, \dodoi{10.1086/301513}

\bibitem[{{Zhang} {et~al.}(2018){Zhang}, {Xu}, {Vasyunin}, {Semenov}, {Wang},
  {Dib}, {Liu}, {Liu}, {Zhang}, {Liu}, {Wang}, {Li}, {Wu}, {Yuan}, {Li}, \&
  {Gao}}]{ZHANG2018}
{Zhang}, G.-Y., {Xu}, J.-L., {Vasyunin}, A.~I., {et~al.} 2018, \aap, 620, A163,
  \dodoi{10.1051/0004-6361/201833622}

\end{thebibliography}
\bibliographystyle{aasjournal}



\end{document}